\documentclass[11pt]{article}

\usepackage[a4paper,margin=1in]{geometry}
\usepackage[utf8]{inputenc}
\usepackage[T1]{fontenc}
\usepackage{lmodern}
\usepackage{microtype}
\usepackage{amsmath,amssymb,amsfonts}
\usepackage{amsthm}
\usepackage{bbm}
\usepackage{booktabs}
\usepackage{multirow}
\usepackage{graphicx}
\usepackage{subfig}
\usepackage{xcolor}
\usepackage{url}
\usepackage[colorlinks=true,linkcolor=blue,citecolor=blue,urlcolor=blue]{hyperref}
\usepackage[round,authoryear]{natbib}
\usepackage{algorithm}
\usepackage[noend]{algpseudocode}
\usepackage{nicefrac}
\usepackage{framed}
\usepackage{caption}

\newtheorem{proposition}{Proposition}

\title{Nonparametric Detection of Multiple Location-Scale Change Points via Wild Binary Segmentation}
\author{Gordon J. Ross\\
Department of Mathematics, University of Edinburgh, United Kingdom\\
\texttt{gordon.ross@ed.ac.uk}}
\date{}

\graphicspath{{img/}}

\begin{document}

\maketitle

\begin{abstract}
Change point methods are used to divide a sequence of observations into
segments with different behaviour. Often, the distributional form of
the observations is unknown, but the changes of interest are likely to involve shifts in location, scale, or both. We consider the problem of
detecting multiple change points in a sequence without specifying a parametric model for the data. We propose the WBS-Lepage procedure, a nonparametric method which combines wild binary
segmentation with a rank-based Lepage statistic. The statistic is formed from Mann--Whitney and Mood components, which are respectively sensitive to changes in location and scale. Since it depends on the observations only through their ranks,
its null distribution is distribution-free. This allows finite-sample thresholds to be calibrated by Monte Carlo simulation,
providing direct control over the probability of falsely detecting change points
when none exist. We compare  WBS-Lepage with existing nonparametric change
point methods, including penalised likelihood and binary-segmentation-based
competitors. The proposed method performs competitively for location changes and
is particularly effective for detecting changes in scale. We illustrate the procedure on a stylometric analysis of changes in an author's writing style and provide an implementation of our method in the accompanying R package \textbf{npwbs}.

\end{abstract}

\section{Introduction}

The multiple change point problem involves the simultaneous estimation of the number and location of $K \geq 0$ change points $\boldsymbol{\tau}= \{\tau_1,\ldots,\tau_K\}$ that partition a sequence $\mathbf{y} = (y_1,\ldots,y_n)$ into $K+1$ segments such that
\[
S_1 = \{y_1,\ldots, y_{\tau_1} \}, \quad
S_2 = \{y_{\tau_1 + 1}, \ldots, y_{\tau_2}\}, \quad
\ldots, \quad
S_{K+1} = \{y_{\tau_K +1},\ldots, y_n\}.
\]
We make the commonly used assumption that the observations are univariate and independent within each segment, so that $y_i \sim_{\mathrm{i.i.d.}} F_j$ if $y_i \in S_j$, for some set of unknown continuous distributions $F_1,\ldots,F_{K+1}$. The continuity assumption is natural for the rank-based procedures considered below, since it avoids ties and ensures that the null distributions of the relevant rank statistics are distribution-free. Our focus in this paper is not on arbitrary changes in the full distribution $F_j$, but on changes that manifest through location and/or scale when the underlying distributional form is otherwise unspecified.

The literature on the single change point problem, where $K \in \{0,1\}$, is vast, with an overview of traditional methods provided by \cite{basseville_detection_1993}. For the multiple change point problem, most existing literature has assumed that the distributional form of each $F_j(\cdot;\theta_j)$ is known, with change points corresponding to shifts in a finite-dimensional parameter $\theta_j$. Recent overviews of parametric techniques for multiple change point detection are provided by \cite{niu_multiple_2016} and \cite{truong_selective_2020}. Broadly, existing methods fall into several categories.

First, sequential approaches such as CUSUM procedures \citep{page_continuous_1954,moustakides_optimal_1986,lai_sequential_1995} and change point model frameworks \citep{hawkins_changepoint_2003} process the sequence one observation at a time and are typically concerned with estimating the location of the most recent change point. Secondly, penalised cost function approaches seek to minimize a criterion of the form
\[
C(\mathbf{y},\boldsymbol{\tau}) + \mathrm{pen}(\boldsymbol{\tau}),
\]
where $C(\mathbf{y},\boldsymbol{\tau})$ is an appropriately chosen cost function, such as a negative log-likelihood, and $\mathrm{pen}(\boldsymbol{\tau})$ penalises the number of change points \citep{hawkins_fitting_2001,killick_optimal_2012,jackson_algorithm_2005,lavielle_using_2005}. Thirdly, stepwise approaches such as binary segmentation transform the multiple change point problem into the task of recursively testing for the existence of a single change point \citep{bai_estimating_1997,cho_multiple-change-point_2015,olshen_circular_2014,vostrikov_detecting_1981,inclan_use_1994,fryzlewicz_wild_2014}. Bayesian approaches, including \cite{barry_bayesian_1993,green_reversible_1995,fearnhead_exact_2006,chib_estimation_1998}, provide another important class of methods, although they take a rather different inferential perspective and will not be our focus here.

In many applications, it is not reasonable to expect the distributional form of the observations within each segment to be known. This has led to the development of nonparametric change point detection algorithms. Much of the classic nonparametric literature focused on the single change point setting \citep{carlstein_nonparametric_1988,pettitt_non-parametric_1979,brodsky_nonparametric_1993,chen_sequential_2019}, although the multiple change point setting is also of considerable interest. Extensions of the sequential approach to the nonparametric setting have been considered by authors such as \cite{hawkins_nonparametric_2010,ross_nonparametric_2011,padilla_sequential_2019}; however, such approaches are not designed for retrospective offline analysis, since inference for a change point at time $t$ is based only on the observations $y_1,\ldots,y_{t-1}$ rather than on the full sequence.

Penalised cost approaches can also be extended to the nonparametric setting, although the absence of a parametric likelihood means that an alternative cost function must be constructed. \cite{zou_nonparametric_2014} proposed a nonparametric likelihood approach to multiple change point detection, and \cite{haynes_computationally_2017} developed a computationally efficient implementation based on the PELT algorithm. This provides a powerful and practically useful class of methods. However, as with other penalised approaches, the finite-sample behaviour of the procedure can be sensitive to the choice of penalty. In particular, default penalties need not provide a desired probability of no-change false detections, so in our empirical comparisons below we consider both default penalties and penalties calibrated to control the false positive probability under the global no-change null.

Binary segmentation methods provide another route to nonparametric multiple change point detection. At each stage of the recursion, the test for a single change point can be carried out using a nonparametric statistic, such as the maximised Mann--Whitney statistic used by \cite{pettitt_non-parametric_1979}. A related approach was taken by \cite{matteson_nonparametric_2014} in a multivariate setting. Such methods have the advantage that the single-segment test can be calibrated under the null hypothesis of no change. However, ordinary binary segmentation can perform poorly when multiple nearby changes mask each other, because it searches greedily for the single best split in the current segment. In some configurations, discussed further in Section~\ref{sec:binarysegmentation}, no single split provides a strong summary of the local changes, even though several changes are visually or statistically apparent.

To address such masking effects, \cite{fryzlewicz_wild_2014} introduced Wild Binary Segmentation (WBS), which maximises a change point statistic over randomly sampled intervals rather than only over the full current segment. By considering local subintervals, WBS can detect changes that may be obscured when the entire sequence is analysed at once. Subsequent random-interval methods have further developed this idea, including Narrowest-Over-Threshold detection \citep{baranowski_narrowest-over-threshold_2019}, WBS2 \citep{fryzlewicz_detecting_2020}, and Isolate-Detect \citep{anastasiou_detecting_2022}. These methods differ in how candidate intervals are generated and selected, but share the broad principle that isolating individual changes within suitable local intervals can substantially improve performance over ordinary binary segmentation.

The present paper develops a rank-based random-interval procedure for detecting multiple location-scale changes in a univariate sequence with unknown continuous distribution. We use ``nonparametric'' in the rank-test sense: the relevant null distributions do not depend on the unknown data-generating distribution. The procedure is not intended to be an omnibus detector of all possible changes in $F_j$; rather, it is designed for changes that are expressed through location and/or scale. This focus is practically important, since changes in mean and variability are among the most common forms of distributional change, and rank-based location-scale tests can have substantially greater power against such alternatives than fully omnibus tests such as Kolmogorov--Smirnov or Cramér--von Mises statistics. Related semiparametric approaches to location and scale change point detection have also recently been considered by \cite{agarwal_semiparametric_2023}. Recent isolation-based methods have also been proposed for fully nonparametric distributional change-point detection \citep{AnastasiouFryzlewicz2025NPID}.

Our method combines a WBS2-style recursive random-interval scheme with a Lepage-type rank statistic formed from standardised Mann--Whitney and Mood components. The Mann--Whitney component is sensitive to changes in location, while the Mood component is sensitive to changes in scale. Since both components depend on the observations only through their ranks, the null distribution of the resulting statistic is distribution-free under the assumption of continuous independent observations. This allows critical values to be precomputed by Monte Carlo simulation and used to provide finite-sample calibration of the probability of detecting a change under the global no-change null. We also give a simple consistency result for the corresponding single change point estimator, showing that the population version of the maximised Lepage statistic is maximised at the true change point whenever either the Mann--Whitney or Mood population contrast is nonzero.

The resulting procedure, which we call WBS-Lepage, is compared empirically with several existing nonparametric change point methods. Our simulations suggest that the method is particularly effective for detecting changes in scale, while remaining competitive for location changes. We also compare against a calibrated version of the nonparametric PELT approach, chosen to have comparable no-change false positive behaviour, in order to separate the effect of the segmentation method from the effect of penalty calibration. A full implementation of the proposed method is provided in the R package \textbf{npwbs}, available from CRAN at \url{https://cran.r-project.org/web/packages/npwbs/index.html}.

The remainder of this paper proceeds as follows. Section~\ref{sec:single} discusses the single change point problem where $K \in \{0,1\}$. Section~\ref{sec:binarysegmentation} reviews binary segmentation and explains why masking can cause difficulties in multiple change point settings. Section~\ref{sec:wbs} introduces the proposed WBS-Lepage procedure, with threshold calibration, pruning, and theoretical properties discussed in Sections~\ref{sec:thresholds}--\ref{sec:theory}. Section~\ref{sec:experiments} compares its performance with existing nonparametric change point methods. Finally, Section~\ref{sec:real} presents an illustrative real data example from stylometry, where the aim is to detect changes in an author's writing style.

\section{Methodology}

\subsection{Detecting a Single Change Point}
\label{sec:single}

We first consider the task of testing for the existence of a single change point $\tau$ at a specific known location $\tau=k$ in a sequence $y_1,\ldots,y_n$ of observations. This can be phrased as a two-sample hypothesis test \citep{basseville_detection_1993}:
\begin{equation}
H_0: y_1,\ldots,y_n \sim_{\mathrm{i.i.d.}} F_0,
\label{eqn:hyp1}
\end{equation}
against
\[
H_1: y_1,\ldots,y_k \sim_{\mathrm{i.i.d.}} F_1,\quad
y_{k+1},\ldots,y_n \sim_{\mathrm{i.i.d.}} F_2,\quad
F_1 \neq F_2.
\]

Let $T^k_{1,n}$ denote a two-sample test statistic chosen with regard to the type of change that we wish to detect. To make the notation clear, we write $T^k_{p,q}$ to denote a two-sample test statistic for the samples
\[
(y_p,y_{p+1},\ldots,y_k)
\quad\text{and}\quad
(y_{k+1},\ldots,y_q).
\]
In the parametric case where the functional forms of $F_0,F_1,F_2$ are known, with only their parameters unknown, a statistic based on the generalized likelihood ratio test is a common choice \citep{basseville_detection_1993}. Now suppose that $k$ is unknown and requires estimation. In this case, the null hypothesis is the same as in \eqref{eqn:hyp1}, but the alternative becomes
\[
H_1: \exists k:\ 
y_1,\ldots,y_k \sim_{\mathrm{i.i.d.}} F_1,\quad
y_{k+1},\ldots,y_n \sim_{\mathrm{i.i.d.}} F_2,\quad
F_1 \neq F_2.
\]

A natural test statistic is obtained by maximizing $T^k_{1,n}$ over all possible split points:
\[
T_{1,n} = \max_{1\leq k<n} |T^k_{1,n}|.
\]
The null hypothesis of no change is rejected if $T_{1,n} > \gamma(n)$ for an appropriately chosen threshold $\gamma(n)$. If the null distribution of $T_{1,n}$ is known, and $\gamma(n)$ is chosen as its $(1-\alpha)$ quantile, then the probability of incorrectly detecting a change under the no-change null is controlled at level $\alpha$.

If the null hypothesis is rejected, an estimate of $\tau$ is given by $\hat{\tau}=\tilde{k}$, where $\tilde{k}$ is the value of $k$ for which $|T^k_{1,n}|$ is maximal. There is a substantial literature on the theoretical properties of such estimators in parametric settings, including asymptotic null distributions and confidence bands for the estimated change point. In the present paper, our focus is instead on the construction and calibration of a rank-based statistic for location-scale changes when the underlying distribution is unknown.

The extension of this testing framework to a nonparametric setting is conceptually straightforward: choose $T^k_{1,n}$ to be a statistic whose null distribution does not depend on the underlying distribution of the observations. One way to achieve this is to use a statistic based on sample ranks \citep{pettitt_non-parametric_1979,brodsky_nonparametric_1993}. In that case, under the assumption of continuous observations, the null distribution of $T_{1,n}$ does not depend on the unknown distribution of the $y_i$'s, and the rejection threshold can be calibrated without specifying a parametric model.

\subsection{Binary Segmentation for Detecting Multiple Change Points}
\label{sec:binarysegmentation}

Next suppose that there are an unknown number $K \geq 0$ of change points $\boldsymbol{\tau}= \{\tau_1,\ldots,\tau_K\}$, and the task is to estimate both $K$ and $\boldsymbol{\tau}$. The single change point procedure described above could in principle be extended by replacing $T_{1,n}$ with a statistic maximized over every possible configuration of change points and over every value $K \in \{1,2,\ldots,n-1\}$. While finding the maximizing configuration can often be achieved efficiently using dynamic programming \citep{killick_optimal_2012}, determining the null distribution of the resulting statistic is substantially more difficult.

This has led to the development of stepwise procedures which avoid direct maximization over all configurations and instead recursively apply tests based on the single change point alternative. Specifically, consider again the hypothesis test
\begin{equation}
H_0: y_1,\ldots,y_n \sim_{\mathrm{i.i.d.}} F_0,
\label{eqn:hyptest2}
\end{equation}
against
\[
H_1: \exists k:\ 
y_1,\ldots,y_k \sim_{\mathrm{i.i.d.}} F_1,\quad
y_{k+1},\ldots,y_n \sim_{\mathrm{i.i.d.}} F_2,\quad
F_1 \neq F_2.
\]
Let
\[
T_{1,n} = \max_{1\leq k<n} |T^k_{1,n}|
\]
for an appropriate statistic $T^k_{1,n}$. The null hypothesis of no change is rejected if $T_{1,n} > \gamma(n)$. Suppose the null hypothesis is rejected and let $\tilde{k}$ be the estimated change point location. The \textbf{binary segmentation} procedure then performs the same test on the observations $y_1,\ldots,y_{\tilde{k}}$ to the left of the detected change point, and on the observations $y_{\tilde{k}+1},\ldots,y_n$ to the right. This process is repeated recursively until no further change points are detected. The final output is an estimate $\hat{K}$ of the number of change points and their locations $\hat{\boldsymbol{\tau}} = (\hat{\tau}_1,\ldots,\hat{\tau}_{\hat{K}})$.

Binary segmentation was introduced by \cite{vostrikov_detecting_1981} and has been widely used since \citep{inclan_use_1994,fryzlewicz_wild_2014}. It is sometimes described as primarily a computational device, since it avoids searching through every possible change point configuration \citep{niu_multiple_2016}. However, another advantage is that it is built from a sequence of single-segment tests. In particular, under the global no-change null, the probability of detecting at least one change point can be calibrated through the threshold used in the initial test.

\begin{figure}[t]
\begin{center}
 \includegraphics[width=0.8\textwidth]{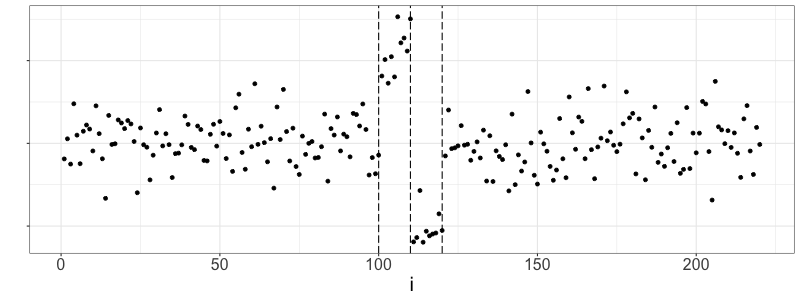}
\end{center}
\caption{Example of a sequence with 3 change points (dotted lines) at locations $\boldsymbol{\tau}=\{100,110,120\}$ where binary segmentation will struggle to detect any changes due to the masking effect.}
\label{fig:wbs}
\end{figure}

Despite these advantages, binary segmentation has a well-known limitation. At each stage in the recursive segmentation process, the segment under consideration is tested against a single change point alternative. This can lead to poor performance when two or more changes mask each other. Consider the sequence shown in Figure~\ref{fig:wbs}. There are three changes in the sequence mean that are visually apparent, but there is no single split point that divides the entire sequence into two segments with substantially different means. As a result, the test in \eqref{eqn:hyptest2} may fail to reject the no-change null, in which case binary segmentation would estimate $\hat{K}=0$. To address this masking problem, \cite{fryzlewicz_wild_2014} introduced Wild Binary Segmentation, which replaces the full-segment statistic with one obtained by maximizing over randomly sampled subintervals. We now adapt this random-interval idea to the rank-based location-scale setting.

\subsection{Nonparametric Wild Binary Segmentation}
\label{sec:wbs}

Our goal is to detect multiple changes in location and/or scale in a univariate sequence of continuous observations, without specifying the underlying distribution. We use a hypothesis testing framework with a rank-based two-sample statistic that is sensitive to both types of change. A natural alternative would be an omnibus statistic such as Kolmogorov--Smirnov or Cramér--von Mises, which can detect general distributional changes. However, such statistics can be underpowered against location-scale alternatives compared with statistics designed specifically for those alternatives. We therefore use a Lepage-type statistic based on standardized Mann--Whitney and Mood components \citep{lepage_combination_1971,ross_nonparametric_2011}.

We first define the statistic on a generic segment $y_p,\ldots,y_q$. Let $p \leq k < q$ be a candidate split point and let
\[
\ell = q-p+1,\qquad n_1 = k-p+1,\qquad n_2 = q-k
\]
denote the segment length and the sizes of the two subsamples. For each $i=p,\ldots,q$, let $r_i^{p,q}$ denote the rank of $y_i$ among $y_p,\ldots,y_q$, so that
\[
r_i^{p,q} = \sum_{j=p}^q I(y_i \geq y_j).
\]
The continuity assumption ensures that ties occur with probability zero, so the ranks have their usual distribution-free null behaviour. For discrete data, one could use midranks or randomized tie-breaking, but we do not pursue this here.

The Mann--Whitney statistic for testing for a location shift at split point $k$ is
\[
U^k_{p,q}
=
\sum_{i=p}^k r_i^{p,q}
-
\frac{n_1(n_1+1)}{2}.
\]
Under the null hypothesis that $y_p,\ldots,y_q$ are identically distributed,
\[
E[U^k_{p,q}] = \frac{n_1 n_2}{2},
\qquad
\operatorname{Var}(U^k_{p,q})
=
\frac{n_1 n_2(\ell+1)}{12}.
\]
The Mann--Whitney statistic is sensitive to changes in location, but it is not generally powerful against pure changes in scale. To capture scale changes, we also use the Mood statistic
\[
M^k_{p,q}
=
\sum_{i=p}^k
\left(
r_i^{p,q} - \frac{\ell+1}{2}
\right)^2.
\]
Under the same no-change null,
\[
E[M^k_{p,q}]
=
\frac{n_1(\ell^2-1)}{12},
\qquad
\operatorname{Var}(M^k_{p,q})
=
\frac{n_1 n_2(\ell+1)(\ell^2-4)}{180}.
\]

We combine these two components into the Lepage-type statistic
\[
L^k_{p,q}
=
\left(
\frac{
U^k_{p,q} - E[U^k_{p,q}]
}{
\sqrt{\operatorname{Var}(U^k_{p,q})}
}
\right)^2
+
\left(
\frac{
M^k_{p,q} - E[M^k_{p,q}]
}{
\sqrt{\operatorname{Var}(M^k_{p,q})}
}
\right)^2.
\]
The original Lepage statistic uses an Ansari--Bradley component for scale \citep{lepage_combination_1971}; our use of the Mood statistic follows earlier work on rank-based monitoring for changes in location and scale \citep{ross_nonparametric_2011}. In our experiments, we found little qualitative difference between the Mood and Ansari--Bradley versions, and use the Mood version throughout.

For a fixed interval $[s,e]$, define the maximized Lepage statistic
\[
T_{s,e}
=
\max_{s\leq k<e} L^k_{s,e}.
\]
Ordinary binary segmentation would compute this statistic only on the full current segment. To reduce the masking problem described above, WBS-type procedures instead compute the statistic over many randomly sampled subintervals.

Let $\mathcal{F}^M_{p,q}$ denote a collection of $M$ random intervals $[s_m,e_m]$ contained within the current segment $[p,q]$, with $p \leq s_m < e_m \leq q$. In practice, intervals shorter than the minimum length required for the calibrated rank test are excluded, as discussed in Section~\ref{sec:thresholds}. The WBS-Lepage statistic on the segment $[p,q]$ is
\begin{equation}
T^{\mathrm{WBS}}_{p,q}
=
\max_{[s_m,e_m]\in \mathcal{F}^M_{p,q}}
T_{s_m,e_m}.
\label{eqn:final}
\end{equation}
The null hypothesis of no change on $[p,q]$ is rejected if
\[
T^{\mathrm{WBS}}_{p,q} > \gamma(q-p+1),
\]
where $\gamma(\cdot)$ is a precomputed threshold depending on the segment length.

If rejection occurs, let
\[
(s_0,e_0)
=
\arg\max_{[s_m,e_m]\in \mathcal{F}^M_{p,q}} T_{s_m,e_m}
\]
be the interval attaining the maximum. The estimated change point within the current segment is then
\[
\tilde{k}
=
\arg\max_{s_0\leq k<e_0} L^k_{s_0,e_0}.
\]
The same procedure is then applied recursively to the two segments $[p,\tilde{k}]$ and $[\tilde{k}+1,q]$.

Our implementation uses the recursive interval-generation strategy of WBS2 \citep{fryzlewicz_detecting_2020}: at each recursive call, new random intervals are generated within the segment currently being analysed. This differs from the original WBS procedure of \cite{fryzlewicz_wild_2014}, in which a fixed collection of intervals is generated at the beginning of the algorithm. We use the term WBS-Lepage for the proposed rank-based random-interval procedure; in the implementation below, the random intervals are regenerated at each recursive call, following the WBS2 convention.

The number of sampled intervals $M$ controls how extensively the current segment is searched. Larger values increase the probability of sampling intervals that isolate individual changes, at the cost of additional computation. In the implementation and experiments below we use $M=10000$ as the default value. We use \(M=10000\) throughout as a conservative computational default, so that
local changes have a high probability of being isolated by at least one sampled
interval; the same value is used for threshold calibration and for all empirical
comparisons.

\subsection{Threshold Determination}
\label{sec:thresholds}

We next discuss the choice of the thresholds $\gamma(n)$ used by the WBS-Lepage procedure. For a segment of length $n$, $\gamma(n)$ is chosen as the $(1-\alpha)$ quantile of the null distribution of $T^{\mathrm{WBS}}_{1,n}$ under the hypothesis that the $n$ observations are independent and identically distributed. Thus, if $\alpha=0.05$, the threshold is chosen so that the probability of detecting a change in a homogeneous segment of length $n$ is approximately $0.05$, up to Monte Carlo error.

Because the Lepage statistic depends on the observations only through their ranks, the null distribution of $T^{\mathrm{WBS}}_{1,n}$ does not depend on the common continuous distribution of the observations. It does depend on the segment length, the interval-sampling scheme, and the chosen value of $M$. This distribution-free property allows the required thresholds to be computed once by Monte Carlo simulation and then reused for any continuous iid null distribution.

Our use of $\gamma(n)$ is a finite-sample calibration for the homogeneity test carried out on the current segment. This should be distinguished from the asymptotic threshold choices used in some parts of the WBS literature, where thresholds are chosen to establish consistency of the full multiple-change-point estimator. Here, when the recursive procedure is applied to a segment $[p,q]$, we use the threshold corresponding to the length of that segment, namely $\gamma(q-p+1)$.

\begin{figure}[t]
\begin{center}
 \includegraphics[width=0.6\textwidth]{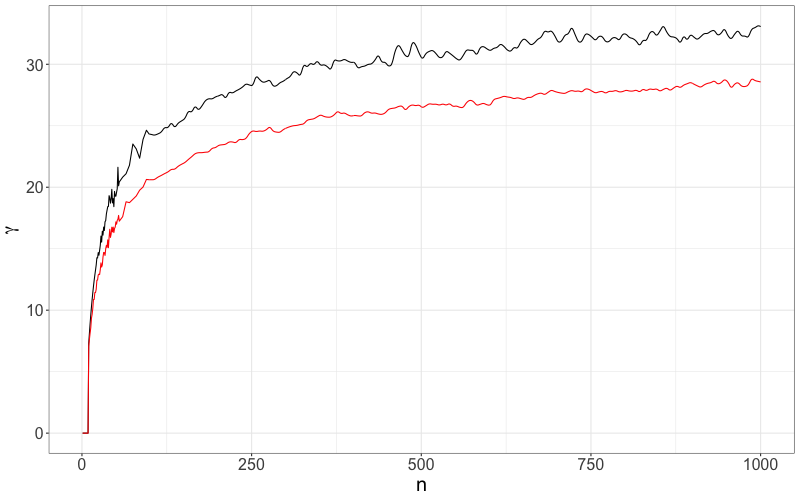}
\end{center}
\caption{Values of the threshold $\gamma(n)$ required to give a false positive probability of 0.05 (red line) and 0.01 (black line) for WBS-Lepage using the Lepage statistic. The plot is shown for $n\leq 1000$ for readability, although thresholds were computed up to $n=10000$.}
\label{fig:thresholds}
\end{figure}

We implemented the threshold computation as follows. For each $n \in \{10,11,\ldots,100\}$, we simulated $1,000$ sequences of length $n$ from a $N(0,1)$ distribution and computed $T^{\mathrm{WBS}}_{1,n}$ for each sequence, using the same interval-sampling scheme and value of $M$ as in the algorithm. For each $n$, the simulated values of $T^{\mathrm{WBS}}_{1,n}$ were ranked, and the relevant empirical percentile was used as the threshold. Since the null distribution is distribution-free, the resulting threshold is valid for any continuous iid distribution, not just the normal distribution used in the simulation.

The same procedure was carried out for $n \in \{105,110,\ldots,995,1000\}$, for $n \in \{1100,1200,\ldots,5000\}$, and for $n \in \{6000,7000,\ldots,10000\}$. Linear interpolation was then used to obtain thresholds for all intermediate values of $n \leq 10000$. Figure~\ref{fig:thresholds} shows the resulting thresholds for $n\leq 1000$ and for significance levels $\alpha=0.05$ and $\alpha=0.01$. Thresholds for both levels are included in the \textbf{npwbs} R package \citep{ross_npwbs_2021}.

The following points should be noted.

\begin{enumerate}
\item Since the procedure detects a change point only when $T^{\mathrm{WBS}}_{p,q}$ exceeds a high null percentile, it may not be possible to detect changes in very short segments. This is a consequence of the discreteness of rank statistics for small sample sizes. For example, with only three observations there are only six possible rank orderings, and hence very few possible values of the maximized statistic. For the Lepage statistic used here, the minimum segment length for which a 5\% calibrated test can both control the false positive probability and still allow rejection is 10. We therefore compute thresholds and sample random intervals only for segments of length at least 10. If true changes occur so frequently that there is insufficient data between consecutive changes to form informative rank comparisons, the method should be expected to underestimate the number of change points.

\item An alternative to precomputing $\gamma(\cdot)$ would be to generate the threshold adaptively during the segmentation procedure using a permutation test, as in the CUSUM context considered by \cite{olshen_circular_2014}. This could allow the threshold to depend on the particular segment being analysed, or allow the nominal level to be adjusted during the recursive procedure. However, doing so would substantially increase the computational cost. We therefore use precomputed thresholds throughout the experiments in Section~\ref{sec:experiments}.

\item At each recursive call, WBS-Lepage generates a new collection of $M$ random intervals within the current segment. This WBS2-style interval generation avoids the depletion of available intervals that can occur if a single fixed collection of intervals is generated only once at the beginning of the algorithm. The proposed method should therefore be understood as a rank-based WBS-style procedure with WBS2-style interval generation.
\end{enumerate}

\begin{algorithm}[t]
\caption{WBS-Lepage}
\begin{algorithmic}[1]

\State Choose a significance level $\alpha$ and number of random intervals $M$
\State Compute or load the corresponding threshold function $\gamma(\cdot)$
\State \textbf{Initialization}: Set $\widehat{\mathcal T} = \textsc{WBS-LP}(\mathbf{y},1,n)$
\State Prune $\widehat{\mathcal T}$ as described in Section~\ref{sec:pruning}
\State Return the pruned set of estimated change points

\Procedure{WBS-LP}{$\mathbf{y},p,q$}
    \If{$q-p+1 < 10$}
        \State Return $\varnothing$
    \EndIf
    \State Generate \(M\) random intervals \(\mathcal F^M_{p,q}\) contained within \([p,q]\), each of length at least 10.
    \State Compute $T^{\mathrm{WBS}}_{p,q}$ from \eqref{eqn:final} using these intervals
    \If{$T^{\mathrm{WBS}}_{p,q} > \gamma(q-p+1)$}
        \State Let $(s_0,e_0) = \arg\max_{[s_m,e_m]\in \mathcal{F}^M_{p,q}} T_{s_m,e_m}$
        \State Let $\tilde{k} = \arg\max_{s_0\leq k<e_0} L^k_{s_0,e_0}$
        \State Return $\textsc{WBS-LP}(\mathbf{y},p,\tilde{k}) \cup \{\tilde{k}\} \cup \textsc{WBS-LP}(\mathbf{y},\tilde{k}+1,q)$
    \Else
        \State Return $\varnothing$
    \EndIf
\EndProcedure

\end{algorithmic}
\label{alg:alg}
\end{algorithm}

\subsection{Pruning}
\label{sec:pruning}

Like other greedy recursive segmentation methods, WBS-Lepage estimates each change point before later changes have necessarily been identified. This can occasionally lead to redundant detections. For example, suppose that the first detected change point is $\tilde{k}_1$, and that subsequent recursive calls detect two further change points $\tilde{k}_0$ and $\tilde{k}_2$ with $\tilde{k}_0 < \tilde{k}_1 < \tilde{k}_2$. It may be that, once $\tilde{k}_0$ and $\tilde{k}_2$ have been identified, the intermediate point $\tilde{k}_1$ is no longer needed: the observations between $\tilde{k}_0$ and $\tilde{k}_2$ may be adequately described as a single homogeneous segment.

We therefore apply a pruning step to remove redundant change points. Similar post-processing ideas have been used in the change point literature, including in \cite{inclan_use_1994} and more recent random-interval methods such as \cite{anastasiou_detecting_2022}. Suppose that the initial WBS-Lepage recursion returns $\hat{K}$ candidate change points
\[
\hat{\tau}_1 < \cdots < \hat{\tau}_{\hat{K}}.
\]
For notational convenience, write $\hat{\tau}_0=0$ and $\hat{\tau}_{\hat{K}+1}=n$. For each candidate change point $\hat{\tau}_j$, $j=1,\ldots,\hat{K}$, we test whether it is needed by considering the combined interval between its neighbouring estimated change points:
\[
a_j = \hat{\tau}_{j-1}+1,\qquad b_j=\hat{\tau}_{j+1}.
\]
On this interval, we test
\[
H_0:
y_{a_j},\ldots,y_{b_j} \sim_{\mathrm{i.i.d.}} F_0
\]
against
\[
H_1:
\exists r \in \{a_j,\ldots,b_j-1\}:
y_{a_j},\ldots,y_r \sim_{\mathrm{i.i.d.}} F_1,\quad
y_{r+1},\ldots,y_{b_j} \sim_{\mathrm{i.i.d.}} F_2,\quad
F_1 \neq F_2.
\]
This test is carried out using the same WBS-Lepage statistic described above. Specifically, we compute $T^{\mathrm{WBS}}_{a_j,b_j}$ using $M$ random intervals contained within $[a_j,b_j]$ and compare it to $\gamma(b_j-a_j+1)$. If the null hypothesis is not rejected, then $\hat{\tau}_j$ is removed from the candidate set.

Since this pruning step can only remove candidate change points and cannot create new ones, it cannot increase the number of detected changes. Its practical effect is assessed in Section~\ref{sec:respruning}.

\subsection{Theoretical properties}
\label{sec:theory}

We record two basic theoretical properties of the proposed statistic. These results justify the null calibration and the single-split estimator used within WBS-Lepage; they are not intended as a full consistency theory for the recursive multiple-change-point procedure.

\begin{proposition}[Distribution-free null calibration]
\label{prop:nullcalibration}
Suppose that $Y_1,\ldots,Y_n$ are independent and identically distributed from a continuous distribution. Fix the interval-sampling scheme and the number of sampled intervals $M$. Then the null distribution of $T^{\mathrm{WBS}}_{1,n}$ depends only on $n$, $M$, and the interval-sampling scheme, and not on the common distribution of the observations. Consequently, if $\gamma_\alpha(n)$ is the $(1-\alpha)$ quantile of this null distribution, then
\[
\Pr\{T^{\mathrm{WBS}}_{1,n} > \gamma_\alpha(n)\}\leq \alpha.
\]
Under the global no-change null, the probability that WBS-Lepage detects at least one change point is therefore at most $\alpha$, up to Monte Carlo error in the approximation of $\gamma_\alpha(n)$.
\end{proposition}

\begin{proof}
Under the null, the rank vector of $Y_1,\ldots,Y_n$ is uniformly distributed over all $n!$ permutations. For fixed sampled intervals, $T^{\mathrm{WBS}}_{1,n}$ is a function only of these ranks and the intervals; averaging over intervals generated independently of the data preserves distribution-freeness. The quantile claim follows directly. Under the global null, the recursive algorithm can return a non-empty set only if the initial full-sequence test rejects.
\end{proof}

Now suppose that there is a single change point:
\[
Y_i\sim
\begin{cases}
F, & i\leq \tau_n,\\
G, & i>\tau_n,
\end{cases}
\qquad
\tau_n/n\to\theta\in(0,1),
\]
where $F$ and $G$ are continuous distributions. Let $H=\theta F+(1-\theta)G$. For independent $X_F\sim F$ and $X_G\sim G$, define
\[
\delta_U=\Pr(X_F>X_G)-\frac12
\]
and
\[
\delta_M=
E\left[\left\{H(X_F)-\frac12\right\}^2\right]
-
E\left[\left\{H(X_G)-\frac12\right\}^2\right].
\]
Here $\delta_U$ is the Mann--Whitney location contrast, while $\delta_M$ measures whether observations from $F$ and $G$ have different dispersion relative to the pooled distribution $H$. The condition $\delta_U^2+\delta_M^2>0$ therefore says that the change is visible to at least one of the two rank components. For fixed $\varepsilon>0$ with $\theta\in(\varepsilon,1-\varepsilon)$, define
\[
\hat\tau_n
=
\arg\max_{\varepsilon n\leq k\leq (1-\varepsilon)n}
L^k_{1,n},
\]
with ties in the maximizer broken arbitrarily.

\begin{proposition}[Single-change-point consistency]
\label{prop:singleconsistency}
Under the single-change-point model above, if $\delta_U^2+\delta_M^2>0$, then
\[
\hat\tau_n/n \xrightarrow{p} \theta .
\]
\end{proposition}

\begin{proof}
We give only a sketch; a fuller derivation of the limiting criterion is given in Appendix~\ref{app:singleproof}. Let $k=\lfloor nt\rfloor$ and write
\[
L^k_{1,n}=Z_U(k)^2+Z_M(k)^2,
\]
where $Z_U$ and $Z_M$ are the standardized Mann--Whitney and Mood components. By uniform convergence of the pooled empirical distribution and standard uniform laws for rank-score partial sums \citep[e.g.][]{shorack_empirical_1986,hajek_theory_1999}, uniformly for $t\in[\varepsilon,1-\varepsilon]$,
\[
\frac{Z_U(\lfloor nt\rfloor)}{\sqrt n}
\xrightarrow{p}
\sqrt{12}\,\delta_U b_\theta(t),
\qquad
\frac{Z_M(\lfloor nt\rfloor)}{\sqrt n}
\xrightarrow{p}
\sqrt{180}\,\delta_M b_\theta(t),
\]
where
\[
b_\theta(t)=
\begin{cases}
(1-\theta)\sqrt{t/(1-t)}, & t\leq\theta,\\[0.8em]
\theta\sqrt{(1-t)/t}, & t\geq\theta.
\end{cases}
\]
Therefore
\[
n^{-1}L^{\lfloor nt\rfloor}_{1,n}
\xrightarrow{p}
\Lambda(t)
=
(12\delta_U^2+180\delta_M^2)b_\theta(t)^2
\]
uniformly on $[\varepsilon,1-\varepsilon]$. Since $\delta_U^2+\delta_M^2>0$, the multiplicative constant is positive. The function $b_\theta(t)^2$ is strictly increasing on $(0,\theta]$ and strictly decreasing on $[\theta,1)$, so $\Lambda(t)$ has a unique maximizer at $t=\theta$. The result follows by the standard argmax theorem.
\end{proof}

Proposition~\ref{prop:singleconsistency} concerns the single-change-point estimator associated with the Lepage statistic. A full consistency result for the recursive WBS-Lepage procedure would require additional assumptions on minimum spacing, signal strength, random interval coverage, and pruning, and is not pursued here. Moreover, the thresholds used below are chosen for finite-sample calibration at a fixed nominal level, rather than for an asymptotic regime in which the nominal level tends to zero. The result is therefore best viewed as a consistency property of the underlying single-split Lepage criterion; the random-interval recursion in Algorithm~1 uses
this criterion locally, but its full multiple-change-point consistency is not claimed here.

\section{Experiments}
\label{sec:experiments}

We now investigate the finite-sample performance of WBS-Lepage for detecting multiple changes in location and/or scale. The aim of these experiments is not to establish uniform superiority over all existing approaches, but to clarify the settings in which the proposed rank-based random-interval method is most useful. In particular, we distinguish between performance on location-change examples, where several existing procedures are strong competitors, and performance on scale-change examples, where a statistic targeted to location and scale changes is expected to have the greatest advantage.

We compare WBS-Lepage with several existing nonparametric change point methods. The first is the divisive partitioning approach of \cite{matteson_nonparametric_2014}, implemented in the \textbf{ecp} R package \citep{james_ecp_2015}. This method is designed for multivariate change point detection, but is also applicable to the univariate examples considered here. The second class of competitors is based on the nonparametric likelihood approach of \cite{zou_nonparametric_2014}, using the computationally efficient PELT implementation of \cite{haynes_computationally_2017} available in the \textbf{changepoint.np} R package \citep{haynes_changepointnp_2021}. In the main simulation tables we report the MBIC default used by this implementation and a calibrated version of the procedure, denoted PELT-FP, whose penalty is chosen to give approximately the same no-change false positive probability as WBS-Lepage. The final competitor is the PYWR method of \cite{padilla_optimal_2021}, which is also based on a WBS-type random-interval idea, but uses different test statistics and thresholding. For WBS-Lepage, we use the pruning step described in Section~\ref{sec:pruning} and thresholds corresponding to $\alpha=0.05$.

\subsection{False Positives}
\label{sec:falsepositives}

In many applications it is not known in advance whether the sequence contains any change points. It is therefore important to understand the behaviour of each method under the global no-change null. WBS-Lepage is calibrated directly for this setting: under iid continuous observations, the threshold construction in Section~\ref{sec:thresholds} controls the probability of detecting at least one change point, up to Monte Carlo error. The ECP method is also based on hypothesis testing and admits a similar calibration. Penalised methods such as PELT do not, under their default penalties, directly target a specified no-change false positive probability. This is not inherently a flaw, but it means that default penalty choices may behave differently from threshold-based procedures in no-change settings.

To study this, we independently simulated 50,000 sequences from each of three distributions: Normal$(0,1)$, Student-$t$ with 3 degrees of freedom, and Lognormal$(1,1/2)$. The latter two distributions represent heavy-tailed and skewed cases respectively. Each simulated sequence consisted of 100 independent observations and contained no change points. We then applied WBS-Lepage, ECP, PYWR, and the nonparametric PELT method with its MBIC default penalty, together with the calibrated PELT-FP version described below. For PYWR we used the default settings supplied by the authors.

\begin{table}[t]
\centering
\begin{tabular}{rrrr}
  \hline
Model  & Normal & Student-t(3) & Lognormal(1,1/2) \\ 
  \hline
WBS-LP  & 0.047 & 0.048 & 0.047 \\ 
ECP&  0.048 & 0.050 & 0.053\\ 
PYWR  & 0.027 & 0.027  & 0.028\\
PELT-MBIC & 0.396 & 0.394 & 0.398  \\ 
PELT-FP & 0.044 & 0.041 & 0.044 \\ 
   \hline
\end{tabular}
\caption{Probability of each change detection method producing at least one false positive detection in a length $100$ sequence of independent observations with no change points.}
\label{tab:falsepos}
\end{table}

Table~\ref{tab:falsepos} shows the proportion of sequences in which at least one change point was detected. WBS-Lepage and ECP both achieve rates close to the nominal level of 0.05, as expected from their threshold calibration. PYWR is conservative in this experiment. By contrast, the default PELT penalties can produce substantially larger no-change detection probabilities, illustrating the sensitivity of penalised methods to the penalty choice when the target is a specified global false positive probability.

To give a fairer comparison with threshold-calibrated methods, we also implemented a calibrated version of PELT, denoted PELT-FP. For a target sequence length $n$, we simulated 10,000 iid $N(0,1)$ sequences of length $n$ and selected the smallest penalty whose empirical probability of detecting at least one change point was no greater than approximately 0.05. This penalty was then used for the corresponding sequence length in the experiments below. The final row of Table~\ref{tab:falsepos} shows that this calibration gives false positive rates close to the nominal level across the three no-change distributions considered here.

\subsection{Change Detection Performance}
\label{sec:changedetection}

We next compare the methods on simulated data sets containing multiple change points. We consider four data models that have previously been used in the change point literature:
\begin{itemize}
  \item \textbf{fms data}, previously studied by \cite{fryzlewicz_wild_2014,frick_multiscale_2014}. This consists of 497 observations with $K=6$ change points at locations
 $\boldsymbol{\tau} = \{139,226,243,300,309,333\}$. The sequence means in each segment are respectively $-0.18,0.08,1.07,-0.53,0.16,-0.69,-0.16$, and the standard deviation $\sigma=0.3$ is constant.
  \item \textbf{mix data}, previously studied by \cite{fryzlewicz_wild_2014}. This consists of 560 observations with $K=13$ change points at locations
  $\boldsymbol{\tau} = \{11,21,41,61,91,121,161,201,251,301,361,421,491\}$. The sequence means in each segment are respectively $7,-7,6,-6,5,-5,4,-4,3,-3,2,-2,1,-1$, and the standard deviation $\sigma=4$ is constant.
  \item \textbf{dhk data}, previously studied by \cite{davies_recursive_2012}. This consists of 1000 observations with $K=9$ change points at locations $\boldsymbol{\tau} = \{100,200,300,400,500,600,700,800,900\}$. The sequence standard deviations in each segment are respectively $2.5,1,2.5,1,2.5,1,2.5,1,2.5,1$, and the sequence mean $\mu=0$ is constant.
  \item \textbf{kfe data}, previously studied by \cite{killick_optimal_2012}. This consists of 1000 observations with $K=5$ change points. Unlike the other data models, these change points are not fixed, and are randomly generated from a uniform distribution on $[30,970]$ with the constraint that there must be at least 30 observations in each segment. The mean of the sequence $\mu=0$ is constant, and the segment standard deviations are randomly generated from a Lognormal$(0,\log(10)/2)$ distribution. The original version of this data model in \cite{killick_optimal_2012} used 10 change points; here we use 5 because the nonparametric segmentation task is substantially more challenging than the parametric version considered there.
\end{itemize}

The first two data models represent changes in location, while the latter two represent changes in scale. We also consider a fifth model designed to represent a short-duration location change inside a longer homogeneous sequence, a setting in which ordinary binary segmentation can struggle because no single full-sequence split is especially natural:
\begin{itemize}
  \item \textbf{interval data}. This consists of 1000 observations with $K=2$ change points at locations
 $\boldsymbol{\tau} = \{490,510\}$. The sequence means in the three segments are respectively $0,2,0$, and the standard deviation $\sigma=1$ is constant.
\end{itemize}

\begin{figure}[t]
\begin{center}
 \subfloat[fms: Normal]{\includegraphics[width=0.27\textwidth]{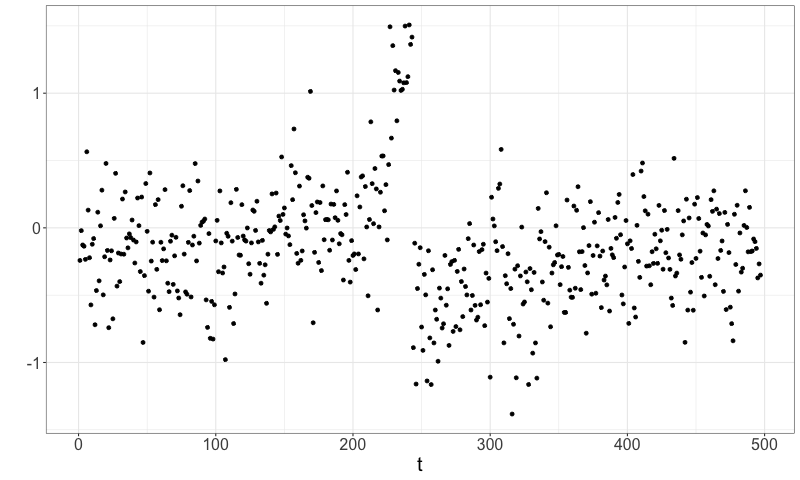}}
  \subfloat[fms: Student-t(3)]{\includegraphics[width=0.27\textwidth]{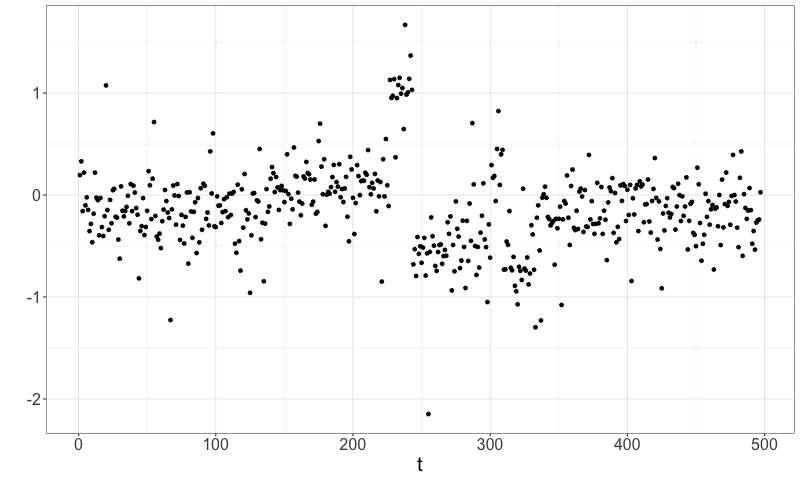}}
 \subfloat[fms: Lognormal(1,1/2)]{\includegraphics[width=0.27\textwidth]{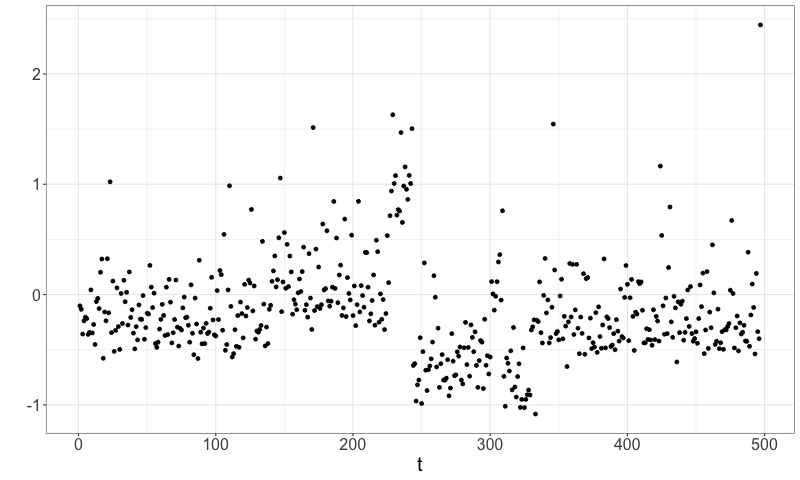}}\\
 \subfloat[mix: Normal]{\includegraphics[width=0.27\textwidth]{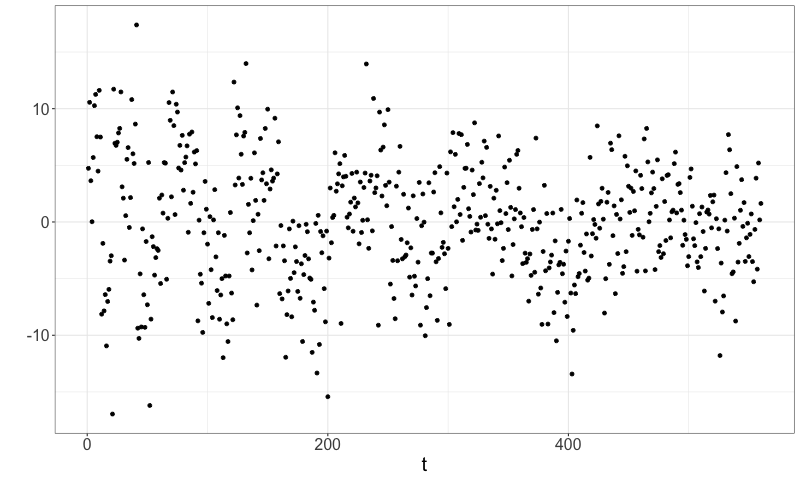}}
  \subfloat[mix: Student-t(3)]{\includegraphics[width=0.27\textwidth]{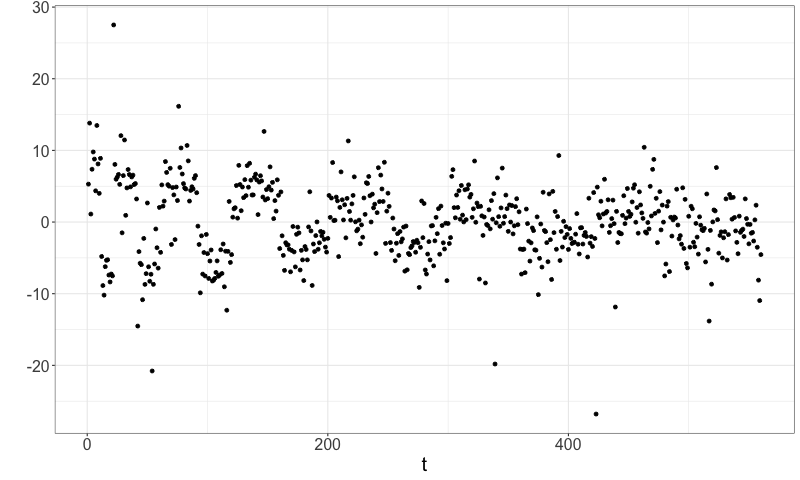}}
 \subfloat[mix: Lognormal(1,1/2)]{\includegraphics[width=0.27\textwidth]{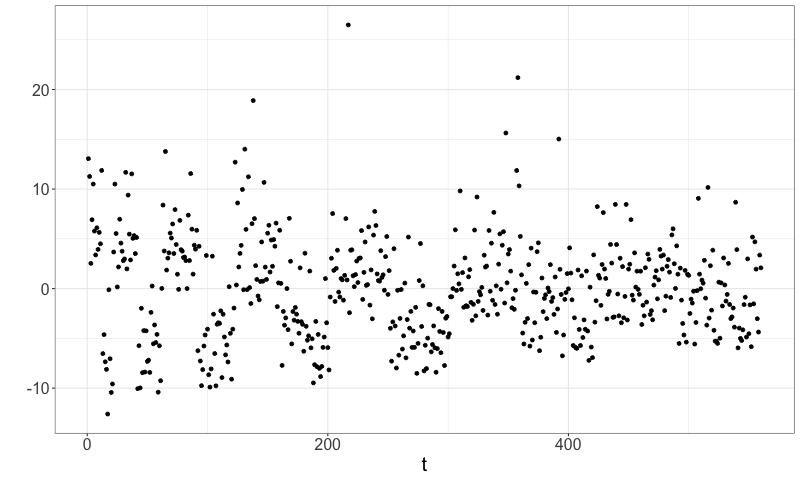}}\\
 \subfloat[interval: Normal]{\includegraphics[width=0.27\textwidth]{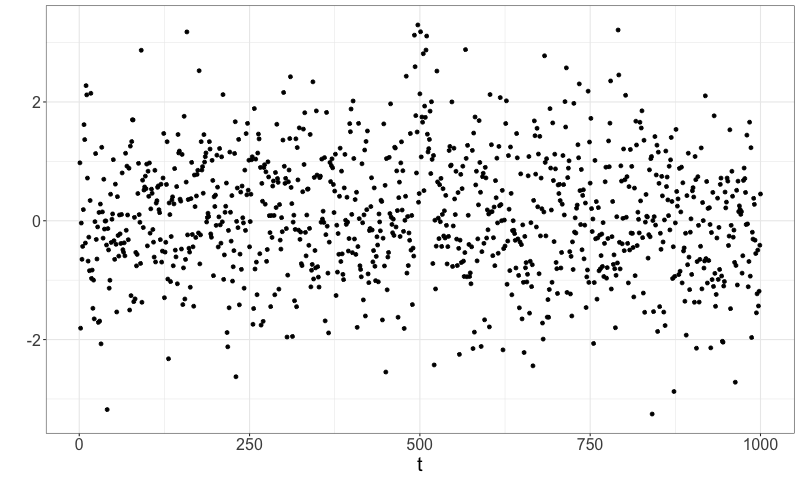}}
  \subfloat[interval: Student-t(3)]{\includegraphics[width=0.27\textwidth]{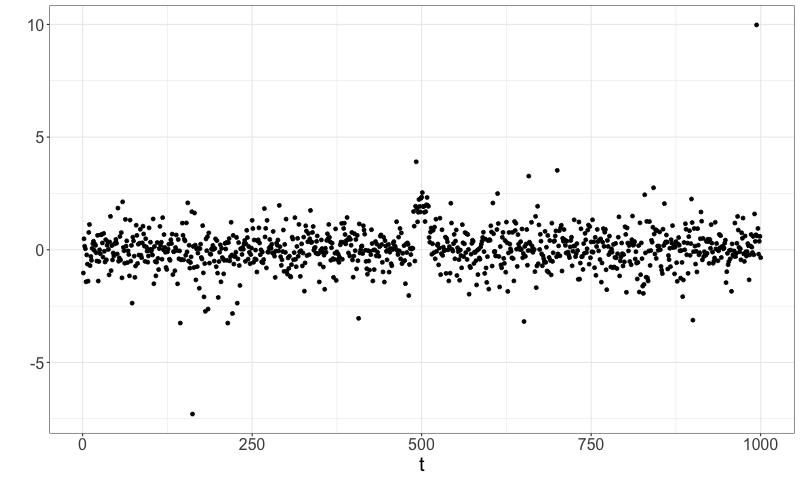}}
 \subfloat[interval: Lognormal(1,1/2)]{\includegraphics[width=0.27\textwidth]{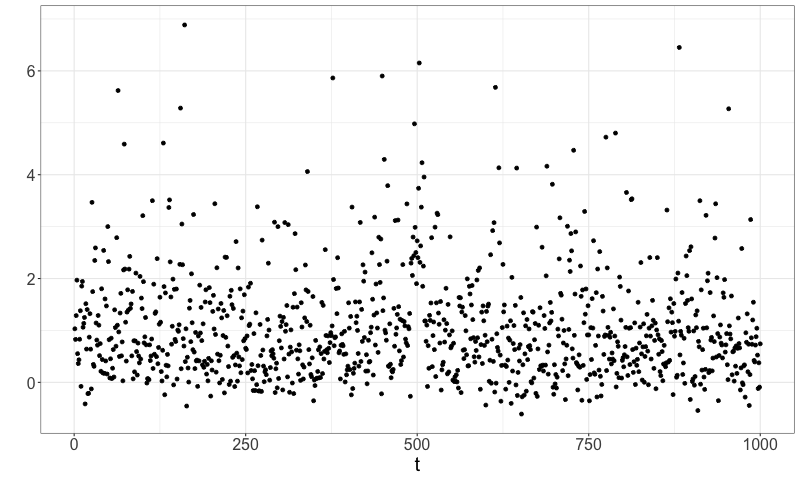}}\\
 \subfloat[dhk: Normal]{\includegraphics[width=0.27\textwidth]{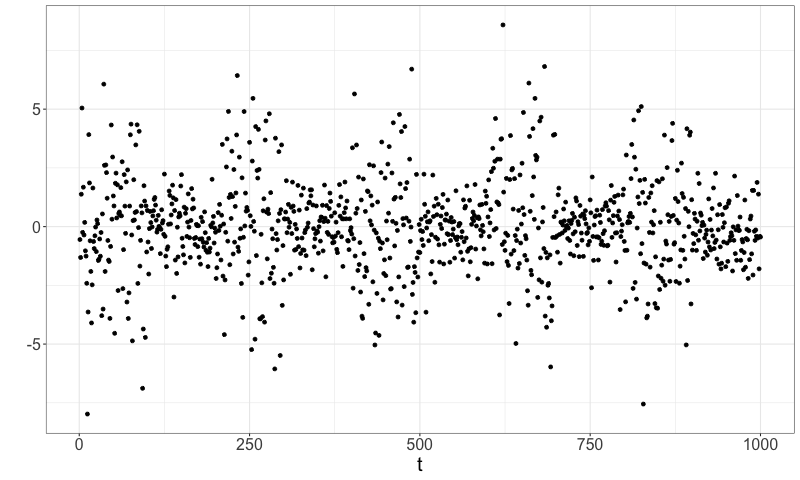}}
  \subfloat[dhk: Student-t(3)]{\includegraphics[width=0.27\textwidth]{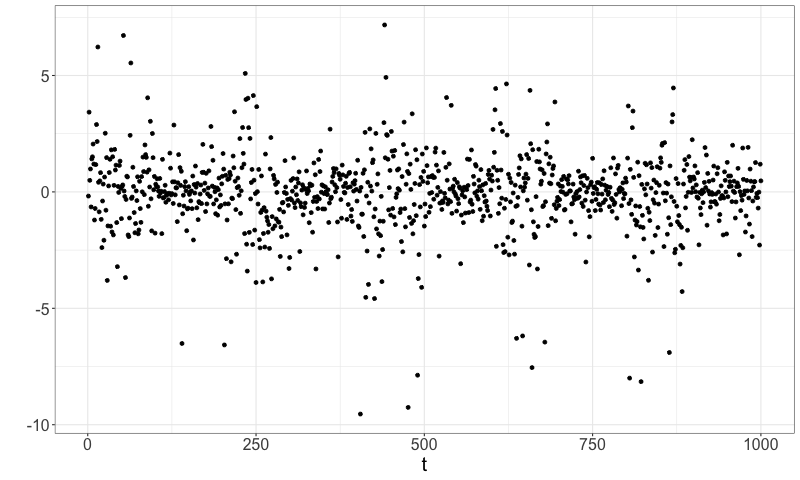}}
 \subfloat[dhk: Lognormal(1,1/2)]{\includegraphics[width=0.27\textwidth]{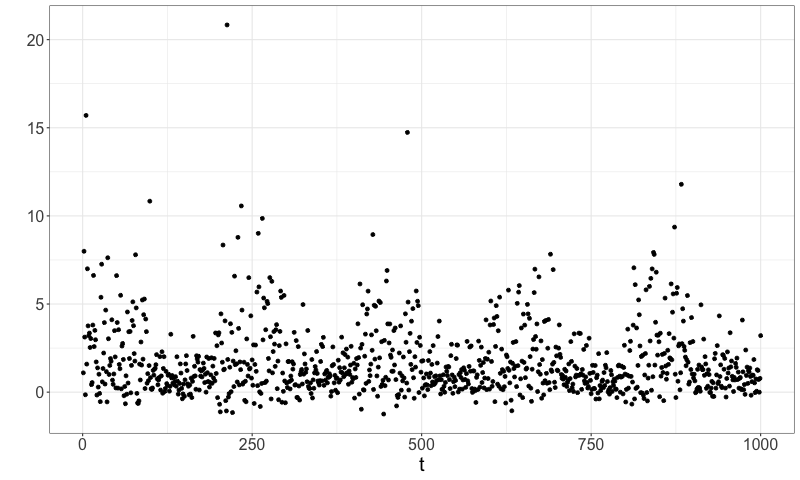}}\\
 \subfloat[kfe: Normal]{\includegraphics[width=0.27\textwidth]{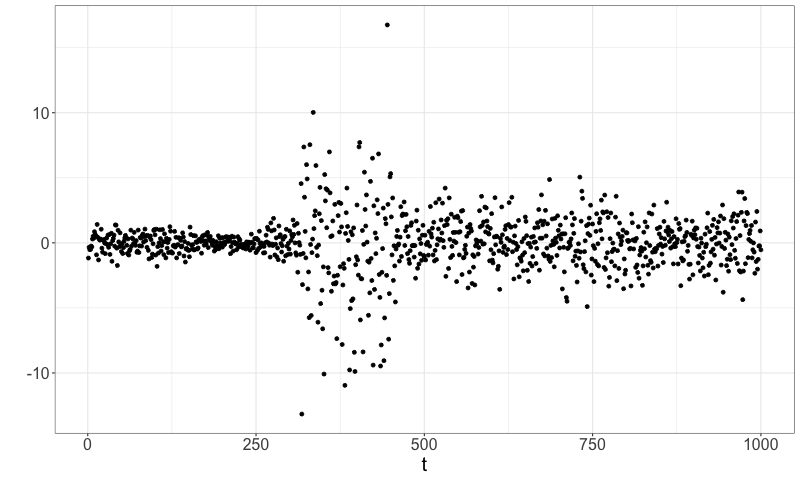}}
  \subfloat[kfe: Student-t(3)]{\includegraphics[width=0.27\textwidth]{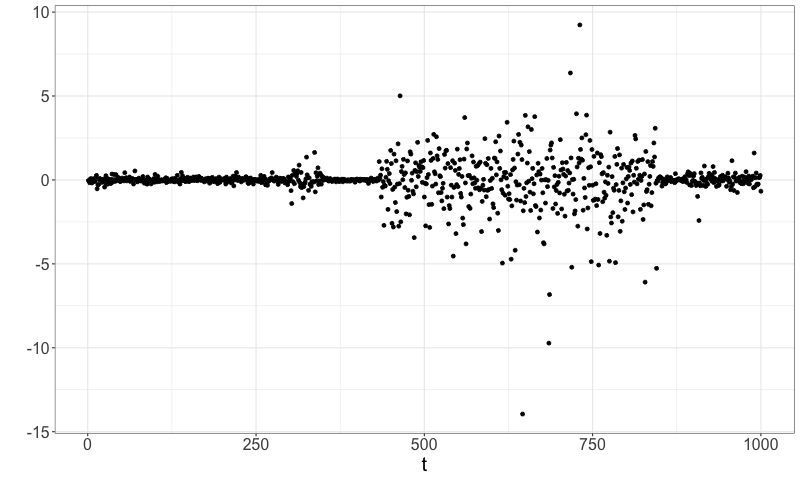}}
 \subfloat[kfe: Lognormal(1,1/2)]{\includegraphics[width=0.27\textwidth]{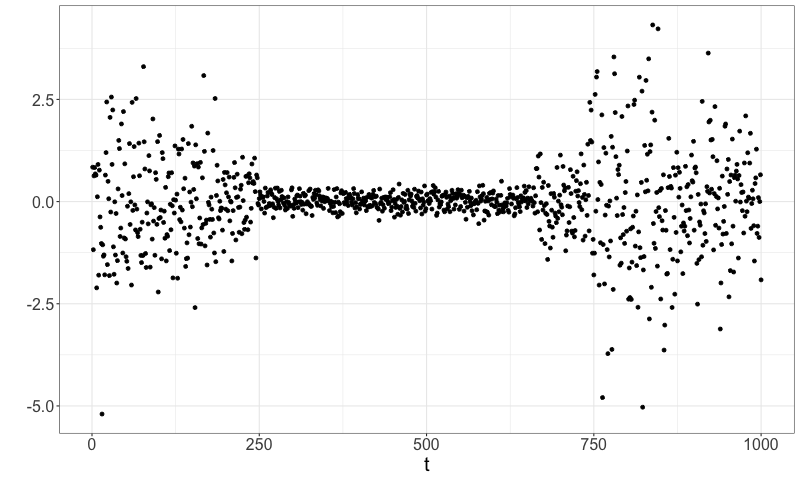}}
\end{center}
\caption{Plots showing a typical sequence simulated from each of the 15 data models. Each row respectively shows the fms, mix, interval, dhk, and kfe data.}
\label{fig:simulateddata}
\end{figure}

For each of the five data models, we consider Normal, Student-$t$ with 3 degrees of freedom, and Lognormal$(1,1/2)$ distributions, giving 15 simulation settings in total. The Student-$t$ and lognormal distributions represent heavy-tailed and skewed cases respectively. In each case the  distribution is centred and scaled to have mean zero and variance one before the segment-specific means and standard deviations above are applied. This ensures that the location examples are location-change examples and the scale examples are scale-change examples under each  distribution. For each setting, 100 independent sequences were simulated. Figure~\ref{fig:simulateddata} shows one typical sequence from each setting.

We assess performance using three summaries. Let $K$ denote the true number of change points and let $\hat{K}_i$ denote the estimated number of change points in the $i$th simulated data set, $i=1,\ldots,100$. Similarly, let $\boldsymbol{\tau}=\{\tau_1,\ldots,\tau_K\}$ denote the true change point locations and let $\hat{\boldsymbol{\tau}}_i=\{\hat{\tau}_{i,1},\ldots,\hat{\tau}_{i,\hat{K}_i}\}$ denote the estimated locations. The first metric is
\[
\frac{1}{100}\sum_{i=1}^{100}|K-\hat{K}_i|,
\]
the average absolute discrepancy in the estimated number of change points. This measures the magnitude of the error in $\hat K$, but not whether a method tends to over- or under-estimate. To address this, Appendix~\ref{app:underover} reports the proportions of simulations in which each method underestimates, exactly estimates, or overestimates $K$.

The second metric is
\[
\frac{1}{100K}\sum_{i=1}^{100}\sum_{k=1}^K I(\tau_k\in\hat{\boldsymbol{\tau}}_i),
\]
the proportion of true change points estimated in their exact locations. Because exact localization can be a stringent requirement in finite samples, we also report the proportion estimated within three observations of the true location,
\[
\frac{1}{100K}\sum_{i=1}^{100}\sum_{k=1}^K
I\left(\exists\hat{\tau}_{i,j}\in\hat{\boldsymbol{\tau}}_i\text{ such that }|\tau_k-\hat{\tau}_{i,j}|\leq 3\right).
\]
The tolerance of three observations is intended only as a small finite-sample localization allowance; exact localization is reported separately.

\begin{table}[t]
\centering
\begin{tabular}{llrrr}
Data & Method & Normal & Student-t(3) & LNorm(1,1/2)\\
  \hline
    \multirow{5}{*}{fms} & WBS-LP & 0.45 & 0.37 & 0.24 \\
    & ECP & \textbf{0.13} & 0.10 & \textbf{0.08} \\
    & PELT-MBIC & 0.16 & 0.46 & 0.18 \\
    & PELT-FP & 0.26 & \textbf{0.06} & 0.09 \\
    & PYWR & 1.57 & 1.67 & 1.57 \\
  \hline
    \multirow{5}{*}{mix} & WBS-LP & 1.26 & 0.79 & 0.78 \\
    & ECP & \textbf{0.82} & \textbf{0.28} & \textbf{0.59} \\
    & PELT-MBIC & 0.99 & 0.88 & 0.77 \\
    & PELT-FP & 2.61 & 2.01 & 2.02 \\
    & PYWR & 5.79 & 4.48 & 4.41 \\
  \hline
    \multirow{5}{*}{interval} & WBS-LP & 0.13 & 0.16 & 0.11 \\
    & ECP & 0.11 & \textbf{0.05} & 0.09 \\
    & PELT-MBIC & 1.08 & 1.79 & 1.73 \\
    & PELT-FP & \textbf{0.03} & \textbf{0.05} & \textbf{0.08} \\
    & PYWR & 2.20 & 2.03 & 2.20 \\
  \hline
    \multirow{5}{*}{dhk} & WBS-LP & \textbf{0.56} & \textbf{0.69} & \textbf{0.54} \\
    & ECP & 6.87 & 7.59 & 6.58 \\
    & PELT-MBIC & 2.28 & 1.94 & 2.17 \\
    & PELT-FP & 1.35 & 7.35 & 0.90 \\
    & PYWR & 8.41 & 8.63 & 6.52 \\
  \hline
    \multirow{5}{*}{kfe} & WBS-LP & \textbf{1.16} & \textbf{1.29} & \textbf{0.91} \\
    & ECP & 2.68 & 2.75 & 2.60 \\
    & PELT-MBIC & 1.67 & 1.55 & 1.65 \\
    & PELT-FP & 1.76 & 1.75 & 1.52 \\
    & PYWR & 3.13 & 3.15 & 2.72 \\
  \hline
\end{tabular}
\caption{Average absolute discrepancy $100^{-1}\sum_{i=1}^{100}|K-\hat{K}_i|$ for each change point method, computed over $100$ realizations of each of the 15 data models. Lower numbers indicate better performance.}
\label{tab:results1}
\end{table}

\begin{table}[t]
\centering
\begin{tabular}{llrrr}
Data & Method & Normal & Student-t(3) & LNorm(1,1/2)\\
  \hline
    \multirow{5}{*}{fms} & WBS-LP & 0.64 & 0.70 & 0.68 \\
    & ECP & 0.58 & 0.64 & 0.60 \\
    & PELT-MBIC & \textbf{0.67} & \textbf{0.74} & \textbf{0.73} \\
    & PELT-FP & 0.65 & 0.73 & 0.72 \\
    & PYWR & 0.14 & 0.16 & 0.14 \\
  \hline
    \multirow{5}{*}{mix} & WBS-LP & 0.50 & 0.58 & 0.59 \\
    & ECP & \textbf{0.56} & \textbf{0.65} & \textbf{0.64} \\
    & PELT-MBIC & 0.51 & 0.58 & 0.61 \\
    & PELT-FP & 0.49 & 0.57 & 0.58 \\
    & PYWR & 0.00 & 0.00 & 0.00 \\
  \hline
    \multirow{5}{*}{interval} & WBS-LP & 0.59 & 0.71 & \textbf{0.73} \\
    & ECP & \textbf{0.62} & \textbf{0.72} & 0.71 \\
    & PELT-MBIC & 0.60 & 0.68 & 0.60 \\
    & PELT-FP & \textbf{0.62} & 0.69 & 0.59 \\
    & PYWR & 0.30 & 0.59 & 0.56 \\
  \hline
    \multirow{5}{*}{dhk} & WBS-LP & \textbf{0.29} & \textbf{0.20} & \textbf{0.35} \\
    & ECP & 0.03 & 0.01 & 0.05 \\
    & PELT-MBIC & 0.26 & 0.15 & 0.27 \\
    & PELT-FP & 0.23 & 0.03 & 0.26 \\
    & PYWR & 0.00 & 0.00 & 0.05 \\
  \hline
    \multirow{5}{*}{kfe} & WBS-LP & \textbf{0.36} & \textbf{0.28} & \textbf{0.37} \\
    & ECP & 0.25 & 0.20 & 0.28 \\
    & PELT-MBIC & 0.33 & 0.25 & 0.33 \\
    & PELT-FP & 0.30 & 0.23 & 0.31 \\
    & PYWR & 0.05 & 0.04 & 0.09 \\
  \hline
\end{tabular}
\caption{Proportion of true change points estimated in their exact locations for each detector, averaged over $100$ realizations of each of the 15 data models. Higher numbers indicate better performance.}
\label{tab:results2}
\end{table}

\begin{table}[t]
\centering
\begin{tabular}{llrrr}
Data & Method & Normal & Student-t(3) & LNorm(1,1/2)\\
  \hline
    \multirow{5}{*}{fms} & WBS-LP & \textbf{0.89} & \textbf{0.96} & 0.92 \\
    & ECP & \textbf{0.89} & \textbf{0.96} & 0.93 \\
    & PELT-MBIC & 0.88 & \textbf{0.96} & \textbf{0.94} \\
    & PELT-FP & 0.85 & 0.95 & 0.93 \\
    & PYWR & 0.55 & 0.64 & 0.60 \\
  \hline
    \multirow{5}{*}{mix} & WBS-LP & 0.77 & 0.87 & 0.86 \\
    & ECP & \textbf{0.82} & \textbf{0.91} & \textbf{0.87} \\
    & PELT-MBIC & 0.80 & 0.83 & \textbf{0.87} \\
    & PELT-FP & 0.74 & 0.78 & 0.80 \\
    & PYWR & 0.41 & 0.54 & 0.53 \\
  \hline
    \multirow{5}{*}{interval} & WBS-LP & 0.94 & 0.98 & \textbf{0.96} \\
    & ECP & \textbf{0.95} & \textbf{0.99} & \textbf{0.96} \\
    & PELT-MBIC & 0.94 & 0.96 & 0.95 \\
    & PELT-FP & \textbf{0.95} & 0.96 & 0.95 \\
    & PYWR & 0.38 & 0.77 & 0.68 \\
  \hline
    \multirow{5}{*}{dhk} & WBS-LP & \textbf{0.71} & \textbf{0.58} & \textbf{0.77} \\
    & ECP & 0.09 & 0.05 & 0.15 \\
    & PELT-MBIC & 0.67 & 0.45 & 0.71 \\
    & PELT-FP & 0.59 & 0.08 & 0.65 \\
    & PYWR & 0.01 & 0.01 & 0.11 \\
  \hline
    \multirow{5}{*}{kfe} & WBS-LP & \textbf{0.61} & \textbf{0.53} & \textbf{0.64} \\
    & ECP & 0.39 & 0.34 & 0.40 \\
    & PELT-MBIC & 0.57 & 0.48 & 0.60 \\
    & PELT-FP & 0.52 & 0.44 & 0.58 \\
    & PYWR & 0.18 & 0.14 & 0.30 \\
  \hline
\end{tabular}
\caption{Proportion of true change points estimated within three observations of their correct locations for each detector, averaged over $100$ realizations of each of the 15 data models. Higher numbers indicate better performance.}
\label{tab:results3}
\end{table}

Tables~\ref{tab:results1}, \ref{tab:results2}, and \ref{tab:results3} show the results. For the location-change examples, ECP and the PELT variants are strong competitors. ECP performs particularly well on the mix data and is competitive on the interval data. PELT-MBIC also localizes well in many location-change settings, although the no-change experiment above shows that this default penalty is not calibrated to the global false positive criterion used here. PELT-FP, by contrast, is explicitly calibrated for that criterion and performs especially well on the interval example, where it nearly always recovers the correct number of change points.

The clearest advantage of WBS-Lepage appears in the scale-change examples. On the dhk and kfe data, WBS-Lepage gives the lowest average number-of-change-point error across all three  distributions. The localization results tell the same broad story: WBS-Lepage is strongest on the scale-change examples under both exact localization and the within-three-observations criterion. ECP and PYWR tend to underestimate the number of variance changes in these settings, while PELT-FP can lose power after calibration, most notably for the heavy-tailed dhk example. PELT-MBIC is more powerful than PELT-FP in some of these scale-change cases, but this must be interpreted together with its no-change behaviour in Table~\ref{tab:falsepos}.

Overall, the proposed method is not uniformly best across all settings. For location changes, ECP and PELT-based methods can match or outperform WBS-Lepage. The main strength of WBS-Lepage in these experiments is its combination of calibrated no-change behaviour, competitive performance for location changes, and consistently strong performance for scale changes across the distributions considered here.

\subsection{The Impact of Pruning}
\label{sec:respruning}

The pruning step in Section~\ref{sec:pruning} is intended to remove redundant change points introduced by the greedy recursive segmentation procedure. It is not a separate methodological contribution, but a post-processing step whose practical effect is worth checking. Table~\ref{tab:pruning} compares the pruned and unpruned versions of WBS-Lepage using the same average absolute discrepancy metric as in Table~\ref{tab:results1}.

\begin{table}[t]
\centering
\begin{tabular}{llrrr}
Data & Method & Normal & Student-t(3) & LNorm(1,1/2)\\
  \hline
    \multirow{2}{*}{fms} & WBS-Unpruned & 0.59 & 0.49 & 0.45 \\
    & WBS-Pruned & \textbf{0.45} & \textbf{0.37} & \textbf{0.24} \\
  \hline
    \multirow{2}{*}{mix} & WBS-Unpruned & \textbf{1.22} & 1.05 & 1.13 \\
    & WBS-Pruned & 1.26 & \textbf{0.79} & \textbf{0.78} \\
  \hline
    \multirow{2}{*}{interval} & WBS-Unpruned & 0.14 & 0.19 & 0.18 \\
    & WBS-Pruned & \textbf{0.13} & \textbf{0.16} & \textbf{0.11} \\
  \hline
    \multirow{2}{*}{dhk} & WBS-Unpruned & 0.99 & 0.92 & 0.74 \\
    & WBS-Pruned & \textbf{0.56} & \textbf{0.69} & \textbf{0.54} \\
  \hline
    \multirow{2}{*}{kfe} & WBS-Unpruned & \textbf{1.10} & \textbf{1.22} & 0.95 \\
    & WBS-Pruned & 1.16 & 1.29 & \textbf{0.91} \\
  \hline
\end{tabular}
\caption{Average absolute discrepancy $100^{-1}\sum_{i=1}^{100}|K-\hat{K}_i|$ for the pruned and unpruned versions of WBS-Lepage across each data set. Lower numbers indicate better performance.}
\label{tab:pruning}
\end{table}

The results show that pruning usually reduces the number-of-change-point error, especially in the fms, interval, and dhk examples. The effect is not uniformly positive in every setting, but the overall pattern supports including the pruning step as the default implementation. When pruning hurts performance, the likely explanation is that a genuine but weak change can be removed when it is retested on the wider interval between neighbouring candidate change points.

\section{Real Data}
\label{sec:real}

We conclude with a real data example drawn from the field of stylometry, where statistical methods are used to answer literary authorship questions. Specifically, we consider the case of Sir Terry Pratchett, a celebrated British author who wrote the Discworld series of fantasy novels, which consists of $41$ books written over a $32$ year period from 1983 to 2015. During 2007, Pratchett was diagnosed with Alzheimer's disease \citep{bbc_news_obituary_2015}, although he continued to write for several years after this diagnosis. In \citep{ross_tracking_2020}, a multivariate parametric change point model was developed to analyse whether there was a detectable change in Pratchett's writing style following this diagnosis. We reanalyse the same corpus in a nonparametric manner.

In stylometry, it is common to represent texts as vectors containing the counts of frequently occurring words. Specifically, let $w_1,\ldots,w_{200}$ denote the $200$ most commonly occurring words across the Discworld corpus. Each book can then be written as a vector $\mathbf{y}_i = (c_{i,1},\ldots,c_{i,200})$ where $c_{i,j}$ denotes the number of times that word $w_j$ appeared in book $i$, with the books arranged in temporal order. A $201^{st}$ element $c_{i,201}$ is also appended to each vector to count the number of words in the book which were not one of the $200$ common words, so that the elements of each vector sum to the total number of words in the corresponding book. For reference, the list of the 200 most common words in the corpus is shown in Figure~\ref{fig:functionwords}, reproduced from \citep{ross_tracking_2020}.

Following \citep{ross_tracking_2020}, we exclude the six young-adult Discworld novels from the formal analysis, since these form a distinct subcorpus and may differ in style for reasons unrelated to chronological stylistic evolution. This leaves $35$ non-young-adult Discworld novels. Each retained book is represented by the $201$-dimensional count vector described above. For the principal-component analysis below, these count vectors are first converted to relative frequencies and the resulting columns are centred and scaled. Thus the principal components are computed from a $35\times 201$ matrix of standardised relative word frequencies, rather than from the raw count matrix.

Given this data representation, \citep{ross_tracking_2020} used a Dirichlet--multinomial change point model to investigate whether there was an abrupt change in the sequence of multivariate word-frequency vectors, while also allowing for gradual drift in writing style. Since WBS-Lepage is univariate, we instead apply it to a low-dimensional summary of the corpus. The first principal component is strongly associated with chronological order and appears to capture the gradual stylistic drift discussed in \citep{ross_tracking_2020}. Since WBS-Lepage is intended to detect abrupt changes rather than smooth stylistic evolution, we use the second principal component as the one-dimensional summary for the change point illustration below. The first principal component is shown as a diagnostic in Appendix Figure~\ref{fig:pratchett-pc1}.

\begin{figure}[t]
\begin{framed}
the, and, to, a, of, I, in, he, was, it, you, that, his, said, her, she, had, with, as, not, but, at, for, on, is, be, have, him, they, all, were, what, me, there, one, my, this, if, no, from, so, would, up, out, by, been, them, or, do, could, when, we, an, are, who, like, know, about, your, did, mr, then, will, very, its, into, their, now, more, man, well, down, which, some, see, back, time, think, just, than, dont, little, can, only, here, any, way, how, over, thought, other, good, say, never, too, looked, much, before, come, two, go, again, old, even, has, made, might, where, head, right, eyes, got, after, mrs, still, yes, hand, off, something, face, should, away, through, must, people, sir, get, though, miss, look, long, us, came, going, went, am, himself, make, why, men, own, big, around, im, those, take, lord, seemed, first, tell, being, always, another, quite, woman, upon, want, things, nothing, last, door, these, didnt, such, oh, knew, once, took, great, really, put, thing, day, young, told, voice, our, let, most, enough, thats, because, every, room, turned, may, left, without, saw, many, course, anything, looking, ever, asked, heard, yet, night, find, done
\end{framed}
\caption{List of the 200 most common words in the Discworld corpus of 41 novels.}
\label{fig:functionwords}
\end{figure}

\begin{figure}[t]
\begin{center}
 \includegraphics[width=0.75\textwidth]{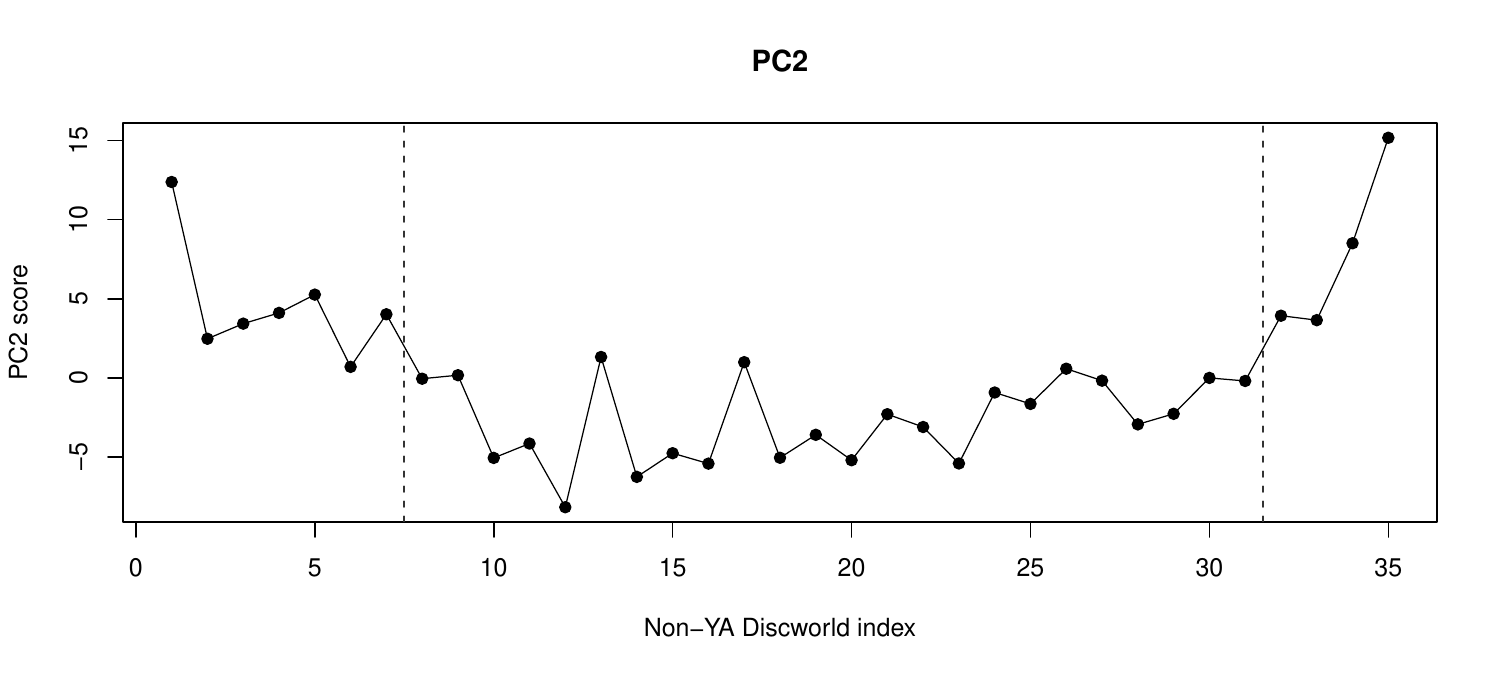}
\end{center}
\caption{Second principal component scores for the 35 non-young-adult Discworld novels, computed from standardised relative frequencies of the 200 most common words plus the ``other words'' component. Dashed vertical lines show the WBS-Lepage change points. A change point at index $k$ denotes a change after book $k$, so lines are drawn between books $k$ and $k+1$. The detected change points occur after non-young-adult books 7 and 31.}
\label{fig:pratchett}
\end{figure}

Applying WBS-Lepage to this sequence of second principal component scores gives two detected change points, after non-young-adult books 7 and 31, shown in Figure~\ref{fig:pratchett}. The first occurs early in the series, around the transition from the earliest Discworld novels to the more established series style. The second occurs between \emph{Thud!} and \emph{Making Money} in the full Discworld ordering; \emph{Wintersmith}, which lies between them chronologically, is excluded from this analysis because it is a young-adult novel. Thus the late change point remains in the same broad period as the abrupt stylistic change identified by \citep{ross_tracking_2020}. This agreement in timing is consistent with the earlier multivariate analysis, but the present analysis is based on an unsupervised one-dimensional projection and should not be interpreted as establishing a causal link with Pratchett's diagnosis.

\section{Conclusion}
The task of detecting multiple change points in a nonparametric setting remains
challenging, particularly when the changes of interest may involve either location
or scale. We have proposed
WBS-Lepage, which combines wild binary
segmentation with a Lepage-type statistic formed from Mann--Whitney and
Mood components. Since the statistic depends on the observations only through
their ranks, its null distribution is distribution-free. This makes it possible to precompute finite-sample thresholds that do not depend on the distributional form of the data.

Our simulation study demonstrates that WBS-Lepage is a strong procedure
for nonparametric location-scale change point detection. On the location-change
examples, it is competitive with established methods such as ECP and
nonparametric PELT, while retaining direct control over the global no-change false positive rate. However, its most distinctive advantage is shown in the scale-change examples, where it gives consistently strong performance across all the distributions considered.

The stylometric example illustrates how WBS-Lepage can be used exploratorily after reducing more complex data to a univariate summary. In that example, the detected late-series change occurs in the same broad period as the previous multivariate Dirichlet--multinomial analysis, while avoiding any strong causal interpretation.

A drawback of our method is the amount of computational time required to simulate from the null distribution of the test statistic in order to generate the $\gamma(n)$ thresholds. This cost is incurred only once, however, since the resulting thresholds are
distribution-free and can be reused across applications. The calibration used here assumes independent observations within homogeneous segments. For serially dependent data, a natural approach is first to model the dependence structure and then apply WBS-Lepage to residuals that are plausibly independent; developing threshold calibrations that explicitly account for serial dependence is a separate problem. We have provided an R implementation in the \textbf{npwbs} package, including
the algorithm, pruning procedure, and precomputed thresholds for the default
settings used above.

\appendix

\section{Proof details for Proposition~\ref{prop:singleconsistency}}
\label{app:singleproof}

We provide the details behind the limiting criterion used in the proof of Proposition~\ref{prop:singleconsistency}. Write $k=\lfloor nt\rfloor$, where $t\in[\varepsilon,1-\varepsilon]$, and let $Z_U(k)$ and $Z_M(k)$ denote the standardized Mann--Whitney and Mood components, respectively:
\[
Z_U(k)
=
\frac{U^k_{1,n}-E_0[U^k_{1,n}]}{\sqrt{\operatorname{Var}_0(U^k_{1,n})}},
\qquad
Z_M(k)
=
\frac{M^k_{1,n}-E_0[M^k_{1,n}]}{\sqrt{\operatorname{Var}_0(M^k_{1,n})}},
\]
where $E_0$ and $\operatorname{Var}_0$ denote the null mean and variance used in the definition of the Lepage statistic.

First consider the Mann--Whitney component. If $t\leq\theta$, then the observations to the left of the candidate split are drawn from $F$, while the observations to the right consist of observations from both $F$ and $G$. Pairs drawn from the same distribution contribute no drift relative to the null expectation, while cross-pairs between $F$ and $G$ contribute $\delta_U$. Hence
\[
E[U^{\lfloor nt\rfloor}_{1,n}]-E_0[U^{\lfloor nt\rfloor}_{1,n}]
=
n^2t(1-\theta)\delta_U+o(n^2).
\]
Since
\[
\operatorname{Var}_0(U^{\lfloor nt\rfloor}_{1,n})
=
\frac{\lfloor nt\rfloor(n-\lfloor nt\rfloor)(n+1)}{12}
=
\frac{n^3t(1-t)}{12}+o(n^3),
\]
it follows that
\[
\frac{E[Z_U(\lfloor nt\rfloor)]}{\sqrt n}
\to
\sqrt{12}\,\delta_U(1-\theta)\sqrt{\frac{t}{1-t}}.
\]
If $t\geq\theta$, the analogous calculation gives
\[
\frac{E[Z_U(\lfloor nt\rfloor)]}{\sqrt n}
\to
\sqrt{12}\,\delta_U\theta\sqrt{\frac{1-t}{t}}.
\]
The stochastic fluctuations around these expectations are $o_p(\sqrt n)$ uniformly over $t\in[\varepsilon,1-\varepsilon]$ by standard uniform laws for rank-score partial sums \citep[e.g.][]{hajek_theory_1999}. Therefore
\[
\frac{Z_U(\lfloor nt\rfloor)}{\sqrt n}
\xrightarrow{p}
\sqrt{12}\,\delta_U b_\theta(t)
\]
uniformly over $t\in[\varepsilon,1-\varepsilon]$.

The Mood component is similar, but we spell out the rank approximation since the statistic is quadratic in the centred ranks. Let $H_n$ be the empirical distribution function of the full sample. Since $\tau_n/n\to\theta$ and $F$ and $G$ are continuous,
\[
\sup_y |H_n(y)-H(y)| \xrightarrow{p}0,
\qquad H=\theta F+(1-\theta)G.
\]
With no ties, $r_i/n=H_n(Y_i)$. Since $u\mapsto (u-1/2)^2$ is Lipschitz on $[0,1]$, replacing $r_i/n$ by $H(Y_i)$ in the leading $n^3$ term of the Mood statistic introduces an $o_p(n^3)$ error, uniformly over $k=\lfloor nt\rfloor$, $t\in[\varepsilon,1-\varepsilon]$ \citep[e.g.][]{shorack_empirical_1986}. Thus the leading term is obtained by averaging $\{H(Y_i)-1/2\}^2$ over the observations to the left of the candidate split. Let
\[
A_F=E\left[\left\{H(X_F)-\frac12\right\}^2\right],
\qquad
A_G=E\left[\left\{H(X_G)-\frac12\right\}^2\right],
\]
so that $\delta_M=A_F-A_G$. Since $H(X)$ is uniform on $(0,1)$ when $X\sim H$,
\[
\theta A_F+(1-\theta)A_G=\frac{1}{12}.
\]
If $t\leq\theta$, the observations to the left of the candidate split are drawn from $F$, and hence
\[
E[M^{\lfloor nt\rfloor}_{1,n}]-E_0[M^{\lfloor nt\rfloor}_{1,n}]
=
n^3t\left(A_F-\frac{1}{12}\right)+o(n^3)
=
n^3t(1-\theta)\delta_M+o(n^3).
\]
Using
\[
\operatorname{Var}_0(M^{\lfloor nt\rfloor}_{1,n})
=
\frac{\lfloor nt\rfloor(n-\lfloor nt\rfloor)(n+1)(n^2-4)}{180}
=
\frac{n^5t(1-t)}{180}+o(n^5),
\]
we obtain
\[
\frac{E[Z_M(\lfloor nt\rfloor)]}{\sqrt n}
\to
\sqrt{180}\,\delta_M(1-\theta)\sqrt{\frac{t}{1-t}}.
\]
For $t\geq\theta$, the same argument gives
\[
\frac{E[Z_M(\lfloor nt\rfloor)]}{\sqrt n}
\to
\sqrt{180}\,\delta_M\theta\sqrt{\frac{1-t}{t}}.
\]
Using the same uniform empirical-distribution and partial-sum arguments,
\[
\frac{Z_M(\lfloor nt\rfloor)}{\sqrt n}
\xrightarrow{p}
\sqrt{180}\,\delta_M b_\theta(t)
\]
uniformly over $t\in[\varepsilon,1-\varepsilon]$.

Combining the two components gives
\[
\frac{1}{n}L^{\lfloor nt\rfloor}_{1,n}
=
\left\{\frac{Z_U(\lfloor nt\rfloor)}{\sqrt n}\right\}^2
+
\left\{\frac{Z_M(\lfloor nt\rfloor)}{\sqrt n}\right\}^2
\xrightarrow{p}
\Lambda(t),
\]
uniformly over $t\in[\varepsilon,1-\varepsilon]$, where
\[
\Lambda(t)=(12\delta_U^2+180\delta_M^2)b_\theta(t)^2.
\]
Since $\delta_U^2+\delta_M^2>0$, the multiplicative constant is positive. Moreover,
\[
b_\theta(t)^2=
\begin{cases}
(1-\theta)^2\dfrac{t}{1-t}, & t\leq\theta,\\[1em]
\theta^2\dfrac{1-t}{t}, & t\geq\theta,
\end{cases}
\]
which is strictly increasing on $(0,\theta]$ and strictly decreasing on $[\theta,1)$. Thus $\Lambda(t)$ has a unique maximum at $t=\theta$, and the stated consistency result follows from the argmax theorem.

\section{Additional simulation diagnostics}
\label{app:underover}

Table~\ref{tab:underover} reports whether the errors in the estimated number of change points in Section~\ref{sec:changedetection} are driven by under- or over-estimation. This complements Table~\ref{tab:results1}, whose absolute-error metric does not distinguish between these two types of error.

\begin{center}
\scriptsize
\resizebox{\textwidth}{!}{%
\begin{tabular}{llccc}
Data & Method & Normal & Student-t(3) & LNorm(1,1/2)\\
  \hline
    \multirow{5}{*}{fms} & WBS-LP & 0.04 / 0.64 / 0.32 & 0.00 / 0.73 / 0.27 & 0.04 / 0.81 / 0.15 \\
    & ECP & 0.01 / 0.89 / 0.10 & 0.01 / 0.93 / 0.06 & 0.02 / 0.94 / 0.04 \\
    & PELT-MBIC & 0.03 / 0.88 / 0.09 & 0.00 / 0.73 / 0.27 & 0.00 / 0.88 / 0.12 \\
    & PELT-FP & 0.14 / 0.86 / 0.00 & 0.03 / 0.97 / 0.00 & 0.04 / 0.94 / 0.02 \\
    & PYWR & 0.53 / 0.24 / 0.23 & 0.62 / 0.16 / 0.22 & 0.55 / 0.20 / 0.25 \\
  \hline
    \multirow{5}{*}{mix} & WBS-LP & 0.57 / 0.25 / 0.18 & 0.31 / 0.41 / 0.28 & 0.25 / 0.45 / 0.30 \\
    & ECP & 0.60 / 0.36 / 0.04 & 0.24 / 0.74 / 0.02 & 0.46 / 0.52 / 0.02 \\
    & PELT-MBIC & 0.53 / 0.37 / 0.10 & 0.24 / 0.36 / 0.40 & 0.41 / 0.40 / 0.19 \\
    & PELT-FP & 0.96 / 0.04 / 0.00 & 0.90 / 0.10 / 0.00 & 0.96 / 0.04 / 0.00 \\
    & PYWR & 0.93 / 0.05 / 0.02 & 0.99 / 0.00 / 0.01 & 0.92 / 0.03 / 0.05 \\
  \hline
    \multirow{5}{*}{interval} & WBS-LP & 0.00 / 0.90 / 0.10 & 0.00 / 0.88 / 0.12 & 0.00 / 0.92 / 0.08 \\
    & ECP & 0.00 / 0.92 / 0.08 & 0.00 / 0.97 / 0.03 & 0.00 / 0.95 / 0.05 \\
    & PELT-MBIC & 0.00 / 0.48 / 0.52 & 0.00 / 0.40 / 0.60 & 0.00 / 0.36 / 0.64 \\
    & PELT-FP & 0.00 / 0.97 / 0.03 & 0.00 / 0.96 / 0.04 & 0.00 / 0.94 / 0.06 \\
    & PYWR & 0.53 / 0.17 / 0.30 & 0.13 / 0.28 / 0.59 & 0.20 / 0.25 / 0.55 \\
  \hline
    \multirow{5}{*}{dhk} & WBS-LP & 0.00 / 0.63 / 0.37 & 0.16 / 0.58 / 0.26 & 0.00 / 0.63 / 0.37 \\
    & ECP & 1.00 / 0.00 / 0.00 & 1.00 / 0.00 / 0.00 & 1.00 / 0.00 / 0.00 \\
    & PELT-MBIC & 0.00 / 0.16 / 0.84 & 0.17 / 0.16 / 0.67 & 0.00 / 0.16 / 0.84 \\
    & PELT-FP & 0.40 / 0.54 / 0.06 & 1.00 / 0.00 / 0.00 & 0.27 / 0.62 / 0.11 \\
    & PYWR & 0.94 / 0.01 / 0.05 & 0.98 / 0.00 / 0.02 & 0.85 / 0.03 / 0.12 \\
  \hline
    \multirow{5}{*}{kfe} & WBS-LP & 0.67 / 0.21 / 0.12 & 0.70 / 0.19 / 0.11 & 0.53 / 0.37 / 0.10 \\
    & ECP & 0.97 / 0.03 / 0.00 & 0.98 / 0.01 / 0.01 & 0.96 / 0.03 / 0.01 \\
    & PELT-MBIC & 0.26 / 0.18 / 0.56 & 0.26 / 0.24 / 0.50 & 0.27 / 0.17 / 0.56 \\
    & PELT-FP & 0.89 / 0.08 / 0.03 & 0.85 / 0.10 / 0.05 & 0.82 / 0.12 / 0.06 \\
    & PYWR & 0.92 / 0.02 / 0.06 & 0.90 / 0.05 / 0.05 & 0.82 / 0.07 / 0.11 \\
  \hline
\end{tabular}%
}
\captionof{table}{Proportions of simulations in which each method underestimated, exactly estimated, or overestimated the number of change points. Each entry is \(P(\hat K < K)/P(\hat K = K)/P(\hat K > K)\).}
\label{tab:underover}
\end{center}

\section{Principal-component diagnostic for the Pratchett example}
\label{app:pratchett-pc1}

Figure~\ref{fig:pratchett-pc1} shows the first principal component scores for the same non-young-adult Discworld corpus used in Section~\ref{sec:real}. This component is dominated by broad chronological drift, and is therefore shown as a diagnostic rather than used as the primary abrupt-change illustration.

\begin{center}
\includegraphics[width=0.75\textwidth]{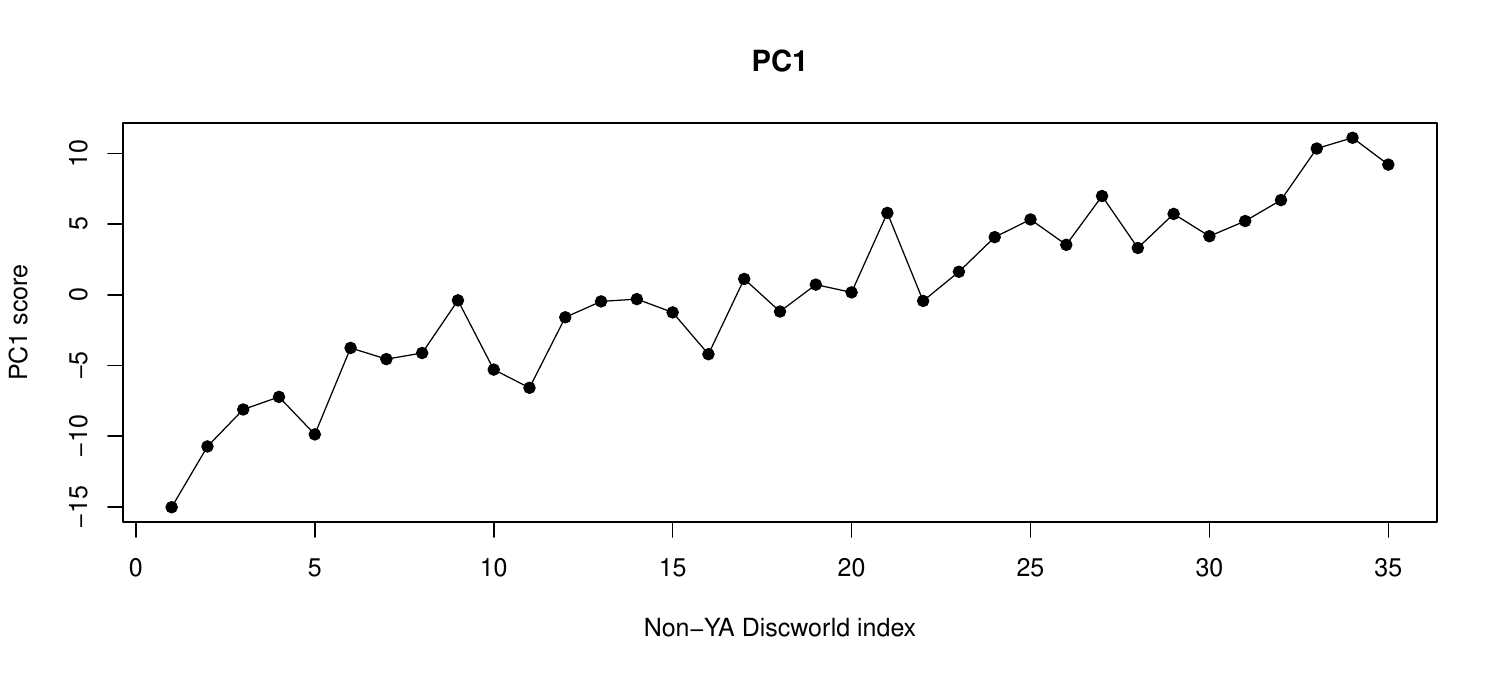}
\captionof{figure}{First principal component scores for the 35 non-young-adult Discworld novels, computed from the same standardised relative word-frequency matrix used in Section~\ref{sec:real}. The first component is dominated by broad chronological drift, and is therefore shown as a diagnostic rather than used as the primary abrupt-change illustration.}
\label{fig:pratchett-pc1}
\end{center}

\bibliographystyle{abbrvnat}
\bibliography{zotero}

\end{document}